\documentclass[12pt]{article}
\usepackage{bbm}
\usepackage{color}
\usepackage{authblk}
\usepackage{latexsym}
\usepackage{epsfig,amssymb,euscript, mathrsfs,cite}
\usepackage{amsmath}
\usepackage{mathrsfs} 
\usepackage{enumerate}
\definecolor{MyDarkBlue}{rgb}{0.15,0.15,0.45}
\usepackage[linktocpage=true]{hyperref}
\hypersetup{
colorlinks=true,
citecolor=MyDarkBlue,
linkcolor=MyDarkBlue,
urlcolor=MyDarkBlue,
pdfauthor={},
pdftitle={},
pdfsubject={hep-th}
}
\newsavebox{\ns}
\newsavebox{\dbrane}
\newsavebox{\dbshort}

\def\be{\begin{equation}}
\def\ee{\end{equation}}
\def\bea{\begin{eqnarray}}
\def\eea{\end{eqnarray}}

\def\v{\vec{v}}
\def\w{\vec{w}}
\def\u{\vec{u}}

\newcommand{\N}{\nonumber}
\newcommand{\nn}{\nonumber}

\newcommand\R{\mathbb{R}}
\newcommand\Z{\mathbb{Z}}

\newcommand\diff{\mathrm{d}}

\newcommand{\dd}{\mathrm{d}}
\newcommand{\me}{\mathrm{e}}
\newcommand{\ii}{\mathrm{i}}

\newcommand{\ex}{\mathrm{e}}
\newcommand{\vol}{\mathrm{vol}}
\newcommand{\Vol}{\mathrm{Vol}}

\newcommand{\betabeta}{B}

\newcommand{\sss}{w}

\newcommand{\cZ}{\mathscr{Z}}
\newcommand{\csugra}{c_{\mathrm{sugra}}}

\newcommand{\BB}{B}

\newcommand{\Ssusy}{S_{\mathrm{SUSY}}}
\newcommand{\pt}{s}
\newcommand{\z}{z}

\newlength{\sswidth}

\numberwithin{equation}{section}       

\begin{document}

\begin{titlepage}

\begin{flushright}
Imperial/TP/2018/JG/04\\
\end{flushright}

\vskip 1.5cm

\begin{center}

{\Large \bf  Toric geometry and the dual of  $c$-extremization}

\vskip 1cm

{Jerome P. Gauntlett$^{\mathrm{a}}$, 
Dario Martelli$^{\mathrm{b},\dagger}$\renewcommand*{\thefootnote}{\fnsymbol{footnote}}
\footnotetext[2]{On leave at the Galileo Galilei Institute, Largo Enrico Fermi, 2, 50125 Firenze, Italy.}and James Sparks$^{\mathrm{c}}$}

\vskip 0.5cm

${}^{\mathrm{a}}$\textit{Blackett Laboratory, Imperial College, \\
Prince Consort Rd., London, SW7 2AZ, U.K.\\}

\vskip 0.2cm
${}^{\mathrm{b}}$\textit{Department of Mathematics, King's College London, \\
The Strand, London, WC2R 2LS,  U.K.\\}

\vskip 0.2 cm
${}^{\,\mathrm{c}}$\textit{Mathematical Institute, University of Oxford,\\
Andrew Wiles Building, Radcliffe Observatory Quarter,\\
Woodstock Road, Oxford, OX2 6GG, U.K.\\}

\end{center}

\vskip 1.5 cm

\begin{abstract}
\noindent  
We consider D3-brane gauge theories at an arbitrary toric Calabi-Yau 3-fold cone singularity that are then further 
compactified on a Riemann surface $\Sigma_g$, with an arbitrary partial topological twist for the global $U(1)$ symmetries. 
This constitutes a rich, infinite class of  two-dimensional $(0,2)$ theories.  Under the assumption that 
such a theory flows to a SCFT, we show that the supergravity formulas for the central charge 
and $R$-charges of BPS baryonic operators of the dual AdS$_3$ solution  may be computed using only the toric data of the Calabi-Yau 3-fold and the topological twist parameters. 
We exemplify the procedure for both the $Y^{p,q}$ and $X^{p,q}$ 3-fold singularities, 
along with their associated dual quiver gauge theories,
showing that the new supergravity results perfectly match 
the field theory results obtained using $c$-extremization, for arbitrary twist over $\Sigma_g$.
We furthermore conjecture that the trial central charge  $\mathscr{Z}$, which we define  in gravity, matches the field theory trial $c$-function off-shell, and show this holds in non-trivial examples. 
Finally, we check our  general geometric formulae against a number of explicitly known supergravity solutions.

\end{abstract}

\end{titlepage}

\pagestyle{plain}
\setcounter{page}{1}
\newcounter{bean}
\baselineskip18pt
\tableofcontents

\newpage

\section{Introduction and summary}\label{sec:intro}

A beautiful feature of SCFTs in $d=2$ spacetime dimensions which preserve $(0,2)$ supersymmetry is that the right moving central charge, $c_R$, can be determined by $c$-extremization \cite{Benini:2012cz,Benini:2013cda}.
One first constructs a trial $R$-symmetry current and an associated trial central charge, proportional to the 't Hooft anomaly for the current, and then extremizes over the space of possible R-symmetries. Generically, this procedure identifies  the exact $R$-symmetry and hence the exact $c_R$. 

In a recent paper\cite{Couzens:2018wnk}  a precise geometric realization of $c$-extremization was formulated for the class of such SCFTs that have a holographic dual in type IIB supergravity of the form AdS$_3\times Y_7$ 
with non-vanishing five-form flux only \cite{Kim:2005ez,Gauntlett:2007ts}. 
In order to set up the geometric version of $c$-extremization it is necessary to take these supergravity solutions off-shell.
This was achieved in \cite{Couzens:2018wnk} by focusing on ``supersymmetric geometries''  in which one imposes the conditions for supersymmetry, {\it  i.e.} the existence of Killing spinors, but relaxes the equation of motion of the five-form. 
The metric on $Y_7$ for these supersymmetric geometries necessarily has a unit norm Killing vector, $\xi$, the ``$R$-symmetry vector field'', which, on-shell, is dual to the $R$-symmetry of the field theory. This Killing vector defines a foliation on $Y_7$, 
and there is an associated six-dimensional transverse K\"ahler metric, satisfying a non-linear PDE and with positive Ricci scalar,
which then determines the full $D=10$ metric and the five-form flux. 
An important feature of the off-shell
supersymmetric geometries is that the eight-dimensional real cone, $C(Y_7)$, over $Y_7$ is a complex cone 
with a non-vanishing holomorphic $(4,0)$-form.

To set up $c$-extremization it was also necessary to impose an additional integral constraint on the supersymmetric geometries, whose precise form we will recall later. The significance of this constraint, 
which is implied by the five-form equation of motion, is that it provides a sufficient condition in order to be able to consistently impose flux quantization of the five-form. Focusing on
this class of supersymmetric geometries, the $c$-extremization begins with a complex cone $C(Y_{7})$,
with a holomorphic $(4,0)$-form, and a holomorphic $U(1)^s$ action. By then choosing a trial 
$R$-symmetry holomorphic vector field $\xi$ together with a transverse K\"ahler metric, with K\"ahler form $J$, one
obtains a supersymmetric geometry. Imposing the integral constraint then allows one to impose flux quantization.
A trial central charge, $\cZ$, can be defined via
\bea\label{cS}
\cZ & \equiv &  \frac{3L^8}{(2\pi)^6g_s^2\ell_s^8}\, \Ssusy(\xi;[J])~,
\eea
where $\Ssusy$ is a supersymmetric action that, importantly, only depends on $\xi$
and the basic cohomology class of $J$. Here $L$ is a length scale which is fixed by flux quantization, and $\ell_s$, $g_s$ are the string length and coupling constant, respectively.
An on-shell supersymmetric geometry then extremizes
 $\cZ$,
and we get $\cZ  |_\mathrm{on-shell}  =  \csugra$, where $\csugra$ is the central charge\footnote{In the context of holographic theories we will simply refer to ``the central charge", when 
we actually mean $c_R$, since the left and right central charges are equal at leading order at large $N$.}
 computed from the supergravity solution, provided it exists, thus completing the identification of a geometric version of $c$-extremization. 

In \cite{Couzens:2018wnk} this formalism was explored in some detail for the special class of examples in which $Y_7=T^2\times Y_5$.
In particular, $Y_5$ is trivially fibred over the $T^2$. Here we will generalize this both by replacing $T^2$ with 
an arbitrary Riemann surface, $\Sigma_g$ of genus $g$ and also allowing $Y_5$ to be non-trivially fibred over 
$\Sigma_g$:
\begin{align}\label{fibration}
Y_5 \ \hookrightarrow \ Y_7 \ \rightarrow  \ \Sigma_g\,.
\end{align} 
From a physical point of view this class of AdS$_3\times Y_7$ solutions can arise
as follows.
If we start with D3-branes sitting at the apex of a Calabi-Yau 3-fold cone singularity, 
we obtain an AdS$_5\times Y_5$ solution of type IIB supergravity with a Sasaki-Einstein metric on $Y_5$.
In the case that $Y_5$ is toric we can extract the dual quiver gauge theory. There is a vast literature on this topic, 
and a variety of approaches, but {\it e.g.} for a recent review on the brane tilings approach, see \cite{Franco:2017jeo}.
 Suppose we then compactify this quiver gauge theory on a Riemann surface $\Sigma_g$ with a ``twist", 
{\it i.e.} also switching on background gauge fields associated with the
$R$-symmetry and other global symmetries, both flavour and baryonic.
In fact, generically ({\it i.e.} for $g\ne 1$), it is essential that we switch on some background fields for
the $R$-symmetry, often called a ``topological twist", in order to preserve
$(0,2)$ supersymmetry. These are the field theories of primary interest in this paper.\footnote{Note, however, that such field theories might not flow to a SCFT in the IR, and examples for the $g=1$ case were discussed
in \cite{Couzens:2018wnk}. In addition, examples of AdS$_3\times Y_7$ 
supergravity solutions were discussed in \cite{Couzens:2018wnk}, again for $g=1$, for which it is unclear how to identify the parent four-dimensional theory. We shall discuss this again in section \ref{secexplexamples}.}
An alternative point of view is to consider D3-branes
wrapping a Riemann surface $\Sigma_g$ inside a Calabi-Yau 4-fold.
 
Assuming that such an AdS$_3\times Y_7$ solution actually exists ({\it i.e.} there are no obstructions), we will show that
the trial central charge in \eqref{cS} can be calculated from a remarkably simple master volume function 
that depends on the toric data of $Y_5$. 
We will give the explicit expression below, after first introducing
the ingredients that enter the formula, most of which are well-known in the context of toric Sasakian geometry.

We start with a toric, K\"ahler cone over $Y_5$, $C(Y_5)$, that is assumed to be Gorenstein, {\it i.e.} have a non-vanishing holomorphic $(3,0)$-form. Recall that the K\"ahler cone metric gives rise to a Sasakian metric on $Y_5$.
Being toric, there are three holomorphic Killing vectors, and the moment maps for the $U(1)^3$ action on $C(Y_5)$
lead to a polyhedral cone, $\mathcal{C}$, with $d\ge 3$ facets which have inward normal vectors $\v_a, \in\mathbb{Z}^3$, $a=1,\dots,d$. A K\"ahler cone metric also specifies a Reeb vector $b_i$, $i=1,2,3$, which lies inside the so-called Reeb cone in $\mathcal{C}^*$. Since the K\"ahler cone is Gorenstein there is a basis in which $\v_a=(1,\w_a)$, with 
$\w_a\in\mathbb{Z}^2$, and this singles out the $b_1$ component. In particular, a Sasaki-Einstein metric on $Y_5$, must have $b_1=3$ \cite{Martelli:2005tp}.

The choice of the Reeb vector $\vec{b}=(b_1,b_2,b_3)$ is associated with a contact one-form, $\eta$, for the Sasakian metric on $Y_5$. Furthermore we have $[\diff\eta]=[\rho]/b_1=2[\omega]$, 
where $\rho$ and $\omega$ are the transverse Ricci-form and K\"ahler form for the Sasakian metric, respectively,
and the cohomology classes refer to the basic cohomology associated with the foliation specified by the Reeb vector.
For our purposes, we are interested in fixing a Reeb vector $\vec{b}$ on the complex cone $C(Y_5)$, maintaining 
$[\diff\eta]=[\rho]/b_1$, but we want to vary the transverse K\"ahler class $[\omega]$. 
As we will recall later, the K\"ahler class is conveniently specified by $d$ parameters $\lambda_a\in\mathbb{R}$, of which $d-2$ give rise to independent 
K\"ahler class parameters. The polytope associated with the K\"ahler cone metric with the given Reeb vector
has $\lambda_a=-\frac{1}{2b_1}$ for all $a=1,\dots,d$. By varying $\lambda_a$ away from these values, we obtain a 
new polytope, and we can use this to obtain the following simple master volume 
formula:
\begin{align}\label{volform}
\mathcal{V}(\vec{b};\{\lambda_a\}) 
& \ \equiv  \ \int_{Y_5}\eta\wedge \frac{1}{2!}\omega^2\,,\\
& \ =\   \frac{(2\pi)^3}{2}\sum_{a=1}^d \lambda_a \frac{\lambda_{a-1}(\v_a,\v_{a+1},\vec{b}) - \lambda_a (\v_{a-1},\v_{a+1},\vec{b})+\lambda_{a+1}(\v_{a-1},\v_a,\vec{b})}{(\v_{a-1},\v_a,\vec{b})(\v_{a},\v_{a+1},\vec{b})}~.\N
\end{align}
Despite appearances, this expression only depends on $d-2$ of the $\lambda_a$.
In the special case that we set $\lambda_a=-\frac{1}{2b_1}$ for all $a=1,\dots,d$, we obtain the formula for 
the volume of a toric Sasakian metric for a given Reeb vector of \cite{Martelli:2005tp}. If we further set
$b_1=3$ and extremize over $b_2,b_3$, we obtain the Reeb vector and the volume for a Sasaki-Einstein metric \cite{Martelli:2005tp}.

Since ${C}(Y_5)$ is toric, there is a $U(1)^3$ action on $Y_5$ and the fibration for $Y_7$ in \eqref{fibration} is specified by three integers $n_i$. In order to ensure that the cone over $Y_7$, $C(Y_7)$,
has a holomorphic $(4,0)$-form, as required for a supersymmetric geometry, there is a corresponding 
restriction on the $n_i$. In a basis for the $U(1)$s in which the holomorphic $(3,0)$-form on the cone over $Y_5$, $C(Y_5)$, has charge 1 under the first $U(1)$ and is uncharged under the second and third,  this restriction is simply that $n_1=2(1-g)$.

Remarkably, the volume formula \eqref{volform} now allows us to easily calculate the central charge 
of the SCFT dual to the AdS$_3\times Y_7$ solution, with $Y_7$ fibred as in \eqref{fibration}
(assuming this solution exists).
In all of the formulae below we need to set $b_1=2$,  {\it after} taking derivatives.
The off-shell supersymmetric action appearing in
the trial central charge, $\cZ$ in \eqref{cS}, is given by
\bea\label{susyactform}
\Ssusy(\vec{b};\{\lambda_a\};A)  &=&  - A \sum_{a=1}^d \frac{\partial \mathcal{V}}{\partial \lambda_a}-4\pi \sum_{i=1}^3 n_i \frac{\partial \mathcal{V}}{\partial b_i}~,
\eea
where the $n_i$ are the integers determining the fibration of $Y_5$ over $\Sigma_g$ with $n_1=2(1-g)$, and $A$ is a parameter that fixes the K\"ahler class of $\Sigma_g$. 
The integral constraint that needs to be imposed on the supersymmetric geometries 
takes the simple form
\bea\label{constraintformulaintro}
0 &=& A\sum_{a,b=1}^d \frac{\partial^2\mathcal{V}}{\partial{\lambda_a}\partial{\lambda_b}}   - 2\pi n_1 \sum_{a=1}^d \frac{\partial \mathcal{V}}{\partial\lambda_a}+4\pi \sum_{a=1}^d\sum_{i=1}^3 n_i \frac{\partial^2 \mathcal{V}}{\partial{\lambda_a}\partial b_i}~.  
\eea
There are two types of five-cycle to consider in imposing flux quantization of the five-form.
Flux quantization over the fibre $Y_5$ at a fixed point on $\Sigma_g$ reads
\bea\label{Nformulaintro}
\frac{2(2\pi \ell_s)^4 g_s}{L^4}N &=& -\sum_{a=1}^d \frac{\partial \mathcal{V}}{\partial\lambda_a}~,
\eea
where $N\in\Z$ and can be interpreted as the number of D3-branes that are wrapping $\Sigma_g$. There
are also flux quantization conditions over the toric three-cycles $S_a$ to be imposed, which read
\bea\label{Maformulaintro}
\frac{2(2\pi \ell_s)^4 g_s}{L^4}M_a &=&  \frac{1}{2\pi}A\sum_{b=1}^d\frac{\partial^2\mathcal{V}}{\partial\lambda_a\partial\lambda_b}+ 2\sum_{i=1}^3 n_i\frac{\partial^2 \mathcal{V}}{\partial\lambda_a\partial b_i}~,
\eea
with $M_a\in\Z$.
The geometric dual of $c$-extremization for this fibred class of $Y_7$ then boils down to
extremizing the supersymmetric action $\Ssusy$, subject to the conditions \eqref{constraintformulaintro}--\eqref{Maformulaintro}, as well as setting $b_1=2$. 
From there one uses \eqref{cS} to obtain
the central charge $\cZ  |_\mathrm{on-shell}  =  \csugra$. We will also show how the
master volume formula can be used to calculate the $R$-charges of a class of baryonic operators in the dual field theory, arising from D3-branes wrapping certain calibrated three-cycles in $Y_7$.

The plan of the rest of the paper is as follows. In section \ref{sec:c} we summarize the geometric
formulation of $c$-extremization for AdS$_3\times Y_7$ solutions of \cite{Couzens:2018wnk}, for general $Y_7$. In section \ref{sec:toric} we derive the master volume formula
\eqref{volform} for $Y_5$ whose cone $C(Y_5)$ is toric and K\"ahler. In section \ref{sec:fibre}
we consider AdS$_3\times Y_7$ solutions with $Y_7$ fibred as in~\eqref{fibration}. We show that
all the ingredients for the geometric dual of $c$-extremization can be derived from the master volume formula, summarising the main results in section~\ref{sec:summary}. 
In section \ref{sec:exist} we briefly discuss the fact that these gravity calculations are valid \emph{provided} the 
AdS$_3\times Y_7$ solution actually exists ({\it i.e.} there are no
obstructions). 
We then turn to some examples.
Section \ref{unitwist} illustrates the formalism for $Y_7$ associated with the so-called universal twist, where the fibration of $Y_5$ over
$\Sigma_g$ is only in the direction of the $R$-symmetry of the Sasaki-Einstein metric on $Y_5$.
Section \ref{secypqdp2} studies the examples for $Y_5$ whose complex cones are given by the $Y^{p,q}$ and $X^{p,q}$ 3-fold singularities. 
These include $Y^{2,1}$ and $X^{2,1}$, which are the canonical complex 
cones over the first and second del Pezzo surfaces, $dP_1$, $dP_2$, respectively.  
The spectacular matching 
between gravity and field theory calculations that we find is, {\it a priori}, a formal matching: in general it assumes 
the supergravity solution exists, and correspondingly the 
field theory results using $c$-extremization are valid provided the theory
flows to the putative SCFT in the IR. 
In section \ref{secexplexamples} we compare our results
with two known classes of explicit supergravity solutions. We conclude with some discussion in
section~\ref{sec:discussion}.


\section{Geometric dual of $c$-extremization}\label{sec:c}

We are interested in supersymmetric AdS$_3$ solutions of type IIB supergravity that are dual to SCFTs with $(0,2)$
supersymmetry, with the ten-dimensional metric and Ramond-Ramond 
self-dual five-form $F_5$ taking the form
\bea
\diff s^2_{10} &=& L^2 \ex^{-B/2}\left(\diff s^2_{\mathrm{AdS}_3} + \diff s^2_{7}\right)~,\N\\
F_5 &=& -L^4\left(\vol_{\mathrm{AdS}_3}\wedge F + *_7 F\right)~.\label{ansatz}
\eea
Here $L$ is an overall dimensionful length scale, with $\diff s^2_{\mathrm{AdS}_3}$ being the metric on a 
unit radius AdS$_3$ with corresponding volume form $\vol_{\mathrm{AdS}_3}$. The 
warp factor $\BB$ is a function on the smooth, compact Riemannian internal space $(Y_7, \diff s^2_7)$ and
$F$ is a closed two-form on $Y_7$ with Hodge dual $*_7 F$. 
In order to define a consistent string theory background we must impose the flux quantization condition
\bea
\frac{1}{(2\pi \ell_s)^4 g_s}\int_{\Sigma_A} F_5 &=& N_A \in \mathbb{Z}~.\label{quantization}
\eea
Here $\ell_s$ is the dimensionful string length,  $g_s$ is the constant string coupling, and
$\Sigma_A\subset Y_7$, with $\{\Sigma_A\}$ forming an integral basis for the free part of $H_5(Y_7,\Z)$. 

The geometric set-up for $c$-extremization in \cite{Couzens:2018wnk} requires that we take these supersymmetric solutions off-shell.
This is achieved by first focusing on {\it supersymmetric geometries}, for which we demand that the ansatz \eqref{ansatz}
still admits the same number of Killing spinors. An on-shell supersymmetric solution
is then obtained by further imposing the five-form equation of motion. The supersymmetric geometries
have a unit norm Killing vector $\xi$, called the $R$-symmetry vector field, which defines 
a foliation $\mathcal{F}_\xi$ of $Y_{7}$. In local coordinates we write $\xi = 2\partial_\z$, and 
the metric takes the form
\bea\label{metric}
\diff s^2_{7} &=& \tfrac{1}{4}(\diff \z + P)^2 + \ex^{\BB} \diff s^2~,
\eea
where $\diff s^2$ is a K\"ahler metric, transverse to the foliation $\mathcal{F}_\xi$. 
The local one-form $P$ is the Ricci one-form of the transverse K\"ahler metric, so that 
$\diff P = \rho$ is the Ricci two-form. The one-form, $\eta$, dual to $\xi$, therefore satisfies
\bea\label{etadef}
\eta &\equiv & \tfrac{1}{2} (\diff\z +P)\,,\qquad \diff\eta \ = \ \tfrac{1}{2}\rho\,.
\eea 
The function $\BB$ in (\ref{metric}) is fixed via $\ex^{\BB} = R/8$,
where $R$ is the Ricci scalar of the transverse K\"ahler metric, and we hence demand that $R>0$.
Finally, the closed two-form  is given by
 \bea\label{fixF}
 F &=& -2J + \diff\left(\ex^{-B}\eta\right)~,
 \eea
 where $J$ is the transverse K\"ahler form. 
 If the orbits of $\xi$ are all closed circles then $Y_7$ is
 called quasi-regular and in the subcase when the circle action is free it is called regular. In these cases $Y_7$ is the total space of a circle bundle over a six-dimensional compact K\"ahler
orbifold or manifold, respectively. If the action of $\xi$ has a non-closed orbit $Y_7$ is said to be irregular.
 
These supersymmetric geometries also solve the five-form equation of motion, and hence become supersymmetric on-shell solutions to type IIB supergravity, provided in addition we 
impose the PDE
\bea\label{boxReqn}
\Box R &=& \tfrac{1}{2}R^2 - R_{ij}R^{ij}~.
\eea
Here $R_{ij}$ denotes the transverse Ricci tensor, and everything in (\ref{boxReqn}) is computed using 
the transverse K\"ahler metric. 
We note that the overall scale of the K\"ahler form $J$ can be absorbed into the 
length scale $L$ in the full AdS$_3$ solution (\ref{ansatz}). 

An important feature of these supersymmetric geometries is that the real cone 
over $Y_{7}$ is a complex cone. More precisely we define the $8$-dimensional cone $C(Y_{7})\equiv \R_{>0}\times Y_{7}$, 
equipped with the conical metric
\bea
\diff s^2_{8} &=& \diff r^2 + r^2 \diff s^2_{7}~.
\eea
While there is no natural K\"ahler structure, or even symplectic structure,  on $C(Y_{7})$, there is a nowhere-zero holomorphic $(4,0)$-form, $\Psi$, which is closed $\diff\Psi=0$, and carries
$R$-symmetry charge two:
\begin{align}\label{chgevf}
\mathcal{L}_\xi\Psi\ = \ 2\ii\Psi\,.
\end{align}
In particular, $\xi$ is a holomorphic vector field.

Putting these supersymmetric geometries
on-shell implies that the action $\Ssusy$ is extremized, where 
\bea\label{Ssusy}
\Ssusy(\xi;[J])  &=& \int_{Y_{7}} \eta\wedge \rho \wedge \frac{1}{2!}J^{2}~.
\eea
It can be shown that for supersymmetric geometries, this is necessarily 
positive: $\Ssusy >0$. 
The action (\ref{Ssusy}) clearly depends on the choice
of $R$-symmetry vector field $\xi$. While it  also appears to depend on both $\rho$ and $J$,
it is not difficult to show that this dependence is only via the
basic cohomology classes $[\rho], [J]\in H^{1,1}_B(\mathcal{F}_\xi)$. Since for fixed complex structure on the cone
 the transverse foliation $\mathcal{F}_\xi$ determines the class
$[\rho]$, we conclude that $\Ssusy$ then only depends on $\xi$ and $[J]$.

We also need to impose flux quantization. To do this, we impose the following additional integral constraint on the supersymmetric geometries:
\bea\label{constraint}
\int_{Y_{7}} \eta \wedge \rho^2 \wedge {J}&=& 0~.
\eea
This is equivalent to the integral of (\ref{boxReqn}) over $Y_7$ holding, which in turn is equivalent to 
imposing an integrated version of the five-form equation of motion. The constraint (\ref{constraint}) also
only depends on the choice of vector field $\xi$ and the basic K\"ahler class $[J]$. Furthermore, it was shown in \cite{Couzens:2018wnk}
that provided we also assume the topological condition $H^2(Y_{7},\R)   \cong  H^2_{{B}}(\mathcal{F}_\xi)/[\rho]$, then \eqref{constraint} is sufficient for consistently imposing flux quantization. In particular this topological condition holds for the fibred geometries (\ref{fibration}) that
we consider in the remainder of the paper, where the fibres $Y_5$ are toric. 
Specifically, 
this condition implies that provided we use representative five-cycles, $\Sigma_A$, on $Y_7$ that are tangent to $\xi$,
then we can 
impose 
\begin{align}
\int_{\Sigma_A}\eta\wedge \rho\wedge J &\ = \ 
 \frac{2(2\pi\ell_s)^4g_s}{L^4}\, N_A\,.
\label{quantize}
\end{align}
In particular, restricting to such cycles, the left hand side only depends 
on the homology class of $\Sigma_A$, as well as $\xi$ and $[J]$. 

We can now summarize the geometric version of $c$-extremization. We fix a complex cone $C(Y_{7})$ 
with holomorphic volume form $\Psi$, and holomorphic $U(1)^s$ action. A general choice of trial $R$-symmetry vector 
may be written as
\bea\label{trialR}
\xi &=& \sum_{i=1}^s b_i\partial_{\varphi_i}\,,
\eea
where $\partial_{\varphi_i}$, $i=1,\ldots,s\geq 1$, are real holomorphic
vector fields generating the $U(1)^{s}$ action on $C(Y_{7})$. For convenience, we choose this basis so that 
the holomorphic volume form has unit charge under $\partial_{\varphi_1}$, but is 
uncharged under $\partial_{\varphi_i}$, $i=2,\ldots,s$, and hence \eqref{chgevf} fixes $b_1=2$.
For a particular choice of $\xi$, and hence foliation 
$\mathcal{F}_\xi$, we then choose a transverse K\"ahler metric with basic class $[J]\in H^{1,1}_{{B}}(\mathcal{F}_\xi)$. 
We impose the constraint \eqref{constraint}, as well as the flux quantization conditions \eqref{quantize}. We then go on-shell by extremizing $\Ssusy$ over the choice of 
$\xi$ and $[J]$ that satisfy the constraints. Equivalently, we extremize the ``trial central charge", $\cZ$, defined by
\bea\label{cS2}
\cZ & \equiv & \frac{3L^8}{(2\pi)^6g_s^2\ell_s^8} \Ssusy~,
\eea
which has the property that for an on-shell supersymmetric solution, {\it i.e.} after extremization,
we obtain the central charge of the dual SCFT:
\bea\label{cS23}
\cZ  |_\mathrm{on-shell} & = & \csugra~.
\eea
An important point to emphasize is that this procedure leads to the central charge associated with a supersymmetric AdS$_3\times Y_7$ solution,
provided that such a supersymmetric solution actually exists. We shall discuss this further 
in  section~\ref{sec:exist}, and also in section \ref{sec:discussion}.

To conclude this section, we note that all known supersymmetric solutions with properly quantized five-form flux are in the regular or quasi-regular class. We expect that this is true in general and
we will also return to this point in section \ref{sec:discussion}.


\section{Toric geometry and the master volume}\label{sec:toric}

As explained in the introduction, in this paper we are interested in $Y_7$ which are fibred over a Riemann surface $\Sigma_g$, so that
\bea
Y_5\ \hookrightarrow\  Y_7\  \rightarrow \ \Sigma_g~.
\eea
In this section we focus on the induced geometry of the fibres $Y_5$, where the $R$-symmetry vector 
$\xi$ is tangent to $Y_5$. The aim is to study the volume function
\bea\label{Vdef}
\mathcal{V} &\equiv & \int_{Y_5}\eta\wedge \frac{1}{2!}\omega^2~.
\eea
Here in a slight abuse of notation $\eta$ is the restriction of (\ref{etadef}) to a fibre, so that 
as in the previous section $\xi\lrcorner\eta=1$, $\xi\lrcorner\diff\eta=0$. Moreover $\omega$ is 
a  transverse K\"ahler form 
for the foliation $\mathcal{F}_\xi$ induced by $\xi$ on $Y_5$. The cone $C(Y_5)=\R_{>0}\times Y_5$ is a complex manifold, 
and again as in section \ref{sec:c} for fixed complex structure $\mathcal{V}=\mathcal{V}(\xi;[\omega])$ 
depends only on the $R$-symmetry vector $\xi$ and the transverse K\"ahler class $[\omega]\in H^2_B(\mathcal{F}_\xi)$. 
In this section\footnote{In later sections we shall also discuss non-convex toric cones, introduced in \cite{Couzens:2018wnk}.} 
 we shall take $C(Y_5)$ to be \emph{toric}, and derive a completely explicit 
formula for $\mathcal{V}$ in terms of toric data, namely (\ref{volform}). We call this the 
``master volume'', because as we shall see in section \ref{sec:fibre}, remarkably everything 
that we need to compute to determine the supergravity formulas for the central charge 
and $R$-charges of BPS baryonic operators may be obtained from $\mathcal{V}$.

\subsection{Toric K\"ahler cones}\label{sec:toricKahler}

Our starting point is to begin with a toric K\"ahler cone in complex dimension $n=3$, as 
first studied in \cite{Martelli:2005tp}. By definition these are K\"ahler metrics 
in real dimension 6 of the conical form
\bea\label{6cone}
\diff s^2_{C(Y_5)} &=& \diff r^2 + r^2\diff s^2_5~,
\eea
which are invariant under a $U(1)^3$ isometry. Introducing generators $\partial_{\varphi_i}$, $i=1,2,3$, 
for each $U(1)$ action, where $\varphi_i$ has period $2\pi$, we also write
\bea\label{Reebbasis}
\xi &=& \sum_{i=1}^3 b_i\partial_{\varphi_i}~.
\eea 
The vector $\vec{b}=(b_1,b_2,b_3)$ parametrizes the choice of $R$-symmetry vector $\xi$.

The complex structure pairs $\xi$ with the radial vector $r\partial_r$, and likewise pairs the dual one-form $\eta$ 
with $\diff r/r$. In particular for K\"ahler cones
\bea\label{detaSasakian}
\diff\eta &=& 2\omega_{\mathrm{Sasakian}}~,
\eea
where $\omega_{\mathrm{Sasakian}}$ is the transverse K\"ahler form. Because $\diff\eta$ 
is also a transverse symplectic form in this case, by definition $\eta$ is a \emph{contact} one-form on $Y_5$. 
The unique vector field $\xi$ satisfying $\xi\lrcorner\eta=1$, $\xi\lrcorner \diff\eta=0$ is then 
also called the \emph{Reeb} vector field. 
We may write the metric on $Y_5$ as
\bea\label{splitmetric}
\diff s^2_5 &=& \eta^2 + \diff s^2_4({\omega})~,
\eea
where $\diff s^2_4({\omega})$ is the transverse K\"ahler metric with K\"ahler form $\omega=\omega_{\mathrm{Sasakian}}$. 
 Moreover, we can define the moment map coordinates
\bea\label{yi}
y_i & \equiv & \tfrac{1}{2}r^2\partial_{\varphi_i}\lrcorner \eta~, \qquad i=1,2,3~.
\eea
These span the so-called moment map polyhedral cone $\mathcal{C}\subset \R^3$, where 
$\vec{y}=(y_1,y_2,y_3)$ are standard coordinates on $\R^3$. The polyhedral cone $\mathcal{C}$, which is convex, may be written as
\bea
\mathcal{C} &=& \{\vec{y}\in \R^3\ \mid \ (\vec{y},\v_a)\geq 0~, \quad a=1,\ldots, d\}~,
\eea
where $\v_a\in\Z^3$ are the inward pointing primitive normals to the facets, and the index $a=1,\ldots,d\geq 3$ labels the facets. 
Geometrically, $\mathcal{C}_{\mathrm{int}}\times U(1)^3$ is a dense open subset of the K\"ahler cone, where 
$\mathcal{C}_{\mathrm{int}}$ denotes the interior of $\mathcal{C}$, with the normal vectors $\v_a\in \Z^3$ to each 
bounding facet in $\partial\mathcal{C}$ specifying which $U(1)\subset U(1)^3$ collapses along that facet. 

An alternative 
presentation is
\bea
\mathcal{C} &=& \Big\{\sum_\alpha t_\alpha \u_\alpha \ \mid \ t_\alpha\geq 0\Big\}~,
\eea
where $\u_\alpha\in\Z^3$ are the outward pointing vectors along each edge of $\mathcal{C}$. Since 
for three-dimensional cones an edge arises as the intersection of two adjacent facets, we may 
 order the facets cyclically around the polyhedral cone, identifying $\v_{d+1}\equiv \v_1$, $\v_0\equiv \v_d$, 
and then note that we may identify the $\alpha$ index, labelling edges, with the $a$ index, labelling 
facets. Specifically, 
\bea\label{vtou}
\u_a &=& \v_{a-1}\wedge \v_a~,
\eea
where $\wedge$ denotes the usual vector cross product in $\R^3$. 
Complex cones $C(Y_5)$ admitting a global holomorphic $(3,0)$-form are called \emph{Gorenstein}, and 
in this case there exists a basis in which $\v_a=(1,\w_a)$, for $\w_a\in\Z^2$. Here 
the holomorphic $(3,0)$-form $\Omega_{(3,0)}$ has unit charge  under $\partial_{\varphi_1}$, and is uncharged under $\partial_{\varphi_2}$, $\partial_{\varphi_3}$ \cite{Martelli:2005tp}. We will henceforth always use such a basis, where we notice that the $b_1$ component of the $R$-symmetry vector $\xi$ in 
(\ref{Reebbasis})
is singled out,  since $\mathcal{L}_\xi \Omega_{(3,0)} = \ii b_1 \Omega_{(3,0)}$.

As shown in \cite{Martelli:2005tp}, for a K\"ahler cone metric the $R$-symmetry vector 
$\vec{b}=(b_1,b_2,b_3)$ necessarily lies in the interior of the Reeb cone, $\vec{b}\in \mathcal{C}^*_{\mathrm{int}}$. 
Here the \emph{Reeb cone} $\mathcal{C}^*$ is by definition the dual cone to $\mathcal{C}$. In particular $\vec{b}\in \mathcal{C}^*_{\mathrm{int}}$
is equivalent to 
 $(\vec{b},\u_a)>0$ for all $a=1,\ldots,d$. 
The Sasakian five-manifold $Y_5$ is embedded at $\{r=1\}$. Using $\xi\lrcorner\eta=1$, together with (\ref{Reebbasis}) and (\ref{yi}), the 
image of $Y_5$ under the moment map is
hence the compact, convex two-dimensional polytope
\bea\label{Ppoly}
P &=& P(\vec{b}) \  \equiv \ \mathcal{C}\cap H(\vec{b})~,
\eea
where the \emph{Reeb hyperplane} is by definition
\bea\label{ReebH}
H &=& H(\vec{b}) \ \equiv \ \Big\{\vec{y}\in \R^3\ \mid \ (\vec{y},\vec{b}) \ = \ \tfrac{1}{2}\Big\}~.
\eea
We will refer to the polytope $P$ in (\ref{Ppoly}) as the \emph{Sasakian polytope}. It sits in the Reeb hyperplane $H(\vec{b})$,
which has normal vector $\vec{b}$.
Notice that the $d$ vertices of the Sasakian polytope $P$ are located at
\bea\label{yaSasakian}
\vec{y}_a &=& \vec{y}_a(\vec{b}) \ \equiv \ \frac{\u_a}{2(\u_a,\vec{b})}~.
\eea
This follows since a vertex is the intersection of the edge $\{t\, \u_a\, \mid \ t\geq 0\}$ with the Reeb hyperplane $H$.

The main object of interest in  \cite{Martelli:2005tp} was the volume of the Sasakian 
manifold $Y_5$. By definition this is
\bea\label{Sasvoldef}
\mathrm{Vol} &\equiv & \int_{Y_5}\eta\wedge \frac{1}{2!}\omega_{\mathrm{Sasakian}}^2~,
\eea
where recall that the transverse K\"ahler form $\omega_{\mathrm{Sasakian}}$ is given by 
(\ref{detaSasakian}). In the following $\mathrm{Vol}$ in (\ref{Sasvoldef}) will always refer to 
the Sasakian volume. 
For Gorenstein K\"ahler cones in fact \cite{Martelli:2006yb}
\bea\label{fixedclass}
[\diff\eta] &=& 2[\omega_{\mathrm{Sasakian}}] \ = \ \frac{1}{b_1}[\rho]\in H^2_B(\mathcal{F}_\xi)~.
\eea
In particular $[\rho]=2\pi c_1^B$, with $c_1^B$ being the basic first Chern class of the foliation. This depends only 
on the complex structure of the cone and the choice of Reeb vector.
The factor of $b_1$
arises since by definition the holomorphic $(3,0)$-form $\Omega_{(3,0)}$ has charge  
$b_1$ under the $R$-symmetry vector $\xi$, so $\mathcal{L}_\xi \Omega_{(3,0)} = \ii b_1 \Omega_{(3,0)}$, as discussed above. Thus we may also write
\bea\label{Sasvolrho}
\mathrm{Vol} &=& \frac{1}{8b_1^2}\int_{Y_5}\eta\wedge \rho^2 ~.
\eea
One of the main results of \cite{Martelli:2005tp} is that the Reeb vector $\xi$ for a 
Sasaki-Einstein metric on $Y_5$ is the unique minimum of $\mathrm{Vol}=\mathrm{Vol}(\vec{b})$, 
subject to the constraint $b_1=3$.
Our master volume  (\ref{Vdef}) is a generalization of the Sasakian volume function, in which we allow for a general transverse 
K\"ahler class $[\omega]$, rather than (\ref{fixedclass}). 
In the remainder of this section 
we derive an equivalent formula for the Sasakian volume function (\ref{Sasvoldef}) that appears in \cite{Martelli:2005tp}, which 
will generalize more readily in the next subsection for any transverse K\"ahler class $[\omega]$. 

In terms of toric geometry, the Sasakian volume (\ref{Sasvoldef}) is \cite{Martelli:2005tp}
\bea\label{SasvolP}
\mathrm{Vol} &=& \mathrm{Vol}(\vec{b}) \ = \ \frac{(2\pi)^3}{|\vec{b}|}\, \vol(P(\vec{b}))~,
\eea
where here $|\vec{b}|=\sqrt{(\vec{b},\vec{b})}$ denotes the Euclidean norm of $\vec{b}$, and $ \vol(P(\vec{b}))$ denotes
the Euclidean area of the Sasakian polytope $P$ defined in (\ref{Ppoly}). In \cite{Martelli:2005tp} a somewhat 
roundabout method was used to compute this, but here we take a direct approach. Specifically, 
 the (signed) area of a compact convex polytope may be obtained by 
choosing any point in the interior, and then summing areas of triangles obtained by joining that interior point 
to each vertex. In turn 
each such triangle area may be written as a two-dimensional 
cross product. Thus pick any point $\vec{y}_0\in P$, and define the area vector
\bea\label{Atriangles}
\vec{A} & \equiv & \frac{1}{2}\sum_{a=1}^d \left(\vec{y}_a-\vec{y}_0\right)\wedge \left(\vec{y}_{a+1}-\vec{y}_0\right)~.
\eea
Here as above $\wedge$ denotes the three-dimensional cross product, and recall we cyclically identify $\vec{y}_{d+1}\equiv \vec{y}_1$.
 Although each term in the sum (\ref{Atriangles}) depends on the choice of $\vec{y}_0$, the area vector 
$\vec{A}$ obtained by summing all contributions does not. 
By construction $\vec{A}$ is orthogonal to the Reeb hyperplane, and hence proportional to the $R$-symmetry vector $\vec{b}$. Provided we choose our cyclic ordering of the $\v_a=(1,\w_a)$ anti-clockwise
in the plane $\R^2\supset \Z^2\ni \w_a$, by the right hand rule the area vector will point in the same direction as the 
 $R$-symmetry vector $\vec{b}$, so that the area of $P$ is 
the inner product
\bea\label{areaproj}
\vol(P(\vec{b})) &=& \Big(\vec{A},\tfrac{\vec{b}}{|\vec{b}|}\Big)~.
\eea
This then gives the Sasakian volume formula
\bea
\mathrm{Vol} &=& (2\pi)^3 \frac{(\vec{A},\vec{b})}{(\vec{b},\vec{b})}\ = \ \frac{\pi^3}{(\vec{b},\vec{b})} \sum_{a=1}^d \Big(\frac{\u_a}{(\u_a,\vec{b})} - 2\vec{y}_0,\frac{\u_{a+1}}{(\u_{a+1},\vec{b})} - 2\vec{y}_0,\vec{b}\Big)~,\label{volnew}
\eea
where $(\cdot,\cdot,\cdot)$ denotes a $3\times 3$ determinant. Notice that although geometrically we required $\vec{y}_0$ to lie in $H$, 
it is straightforward to see that 
(\ref{volnew}) is completely independent of $\vec{y}_0\in\R^3$. In particular we may choose to set $\vec{y}_0=0$. 
In fact for Gorenstein K\"ahler cones there is in some sense a more natural (non-zero) choice of 
 $\vec{y}_0$, as we shall see shortly. The formula (\ref{volnew}) is at first sight different to 
that appearing in \cite{Martelli:2005tp}, but of course by construction it must be equivalent. We shall explicitly 
recover the formula appearing in \cite{Martelli:2005tp} in the next subsection, after first generalizing 
(\ref{volnew}).

We conclude this subsection by noting that although we have so far phrased everything in terms of toric Sasakian geometry, 
in fact the volume function (\ref{volnew}) has more general validity. In particular, for 
the application to the off-shell AdS$_3$ geometries, recall that the cone $C(Y_7)=\R_{>0}\times Y_7$ is 
complex but in fact not K\"ahler. For $Y_7$ of the fibred form (\ref{fibration}), the cones over the fibres
$C(Y_5)=\R_{>0}\times Y_5$ are complex, and by assumption also toric, 
 with the $R$-symmetry vector written as in (\ref{Reebbasis}) in terms of the holomorphic $U(1)^3$ action.
The formula (\ref{Sasvolrho}) immediately implies that 
the Sasakian volume is an invariant of a complex cone, together with 
a choice of $R$-symmetry vector. This fact also follows from the relation of 
 the Sasakian volume to the index-character/Hilbert series derived  in \cite{Martelli:2006yb}, since the 
latter depend only on the complex geometry of the cone, and not on a K\"ahler structure.
In fact it's the formula (\ref{Sasvolrho}) for the Sasakian volume that is directly relevant to our problem, 
and which we shall generalize next.

\subsection{Varying the transverse K\"ahler class}\label{sec:vary}

In this section we would like to derive an analogous formula to (\ref{volnew}) for the 
master volume function (\ref{Vdef}). In fact the Sasakian volume (\ref{volnew}) is the master 
volume in the special case that 
the transverse K\"ahler class 
is $[\omega]=[\omega_{\mathrm{Sasakian}}]=[\rho]/2b_1$. Thus, 
as in the previous subsection, we fix a Gorenstein toric  complex cone $C(Y_5)$ and also choose an $R$-symmetry vector $\vec{b}$ as in (\ref{Reebbasis}). 
We would then like to vary the transverse K\"ahler class. This is perhaps easiest to think
about first in the quasi-regular case, 
where by definition the $R$-symmetry vector generates a $U(1)$ action and $V\equiv Y_5/U(1)$ is a K\"ahler orbifold. Then $H^2_B(\mathcal{F}_\xi)\cong H^2(V;\R)$, and for a K\"ahler cone metric on $C(Y_5)$ the K\"ahler class of the K\"ahler form $\omega_{\mathrm{Sasakian}}$ 
on this base is fixed to be proportional to the topological class $[\rho]$, as in (\ref{fixedclass}). In particular, varying the 
K\"ahler class $[\omega]$ away from $\omega_{\mathrm{Sasakian}}$ means that the metric (\ref{splitmetric}) on $Y_5$ will no longer be Sasakian.
On the other hand, changing the K\"ahler class is well understood in toric geometry for the base $V$: it 
 simply moves the edge vectors of the polytope $P$ parallel to themselves. We begin by reviewing this, following  \cite{guillemin1994}, before generalizing the discussion 
to the case at hand. 

Let $V$ be a  compact toric K\"ahler four-manifold $V$. There is an associated moment map, with moment map image being 
a compact convex polytope  $\Delta\subset \R^2$, with 
inward pointing primitive edge vectors $\vec{n}_a\in\Z^2$, $a=1,\ldots,d$. The polytope may then be written as
\bea\label{Delta2d}
\Delta &=& \{\vec{x}\in \R^2 \ \mid \ (\vec{x},\vec{n}_a) \geq \lambda_a~, \quad a=1,\ldots, d\}~.
\eea
The parameters $\lambda_a\in\R$ determine the K\"ahler class. Explicitly \cite{guillemin1994}
\bea\label{omegalambda}
[\omega] &=& -2\pi\sum_{a=1}^d \lambda_a c_a \in H^2(V;\R)~,
\eea
where $c_a\in H^2(V;\Z)$ are Poincar\'e dual to the $d$ toric divisors. By definition the latter are the torus-invariant divisors, 
which map to 
the edges of the polytope $\Delta$ under the moment map. Since $\dim H^{1,1}(V,\R)=d-2$, in fact 
the K\"ahler class itself depends on only
$d-2$ of the $d$ parameters $\{\lambda_a\}$. 
We also recall that
\bea\label{rhoca}
[\rho] &=& 2\pi\sum_{a=1}^d c_a  \in H^2(V;\R)~.
\eea

We would now like to apply the above formalism of  \cite{guillemin1994} in the transverse setting, where $\R^2$ is identified with the Reeb hyperplane 
$H$ and the moment map polytope $\Delta\subset\R^2$ is identified with the Sasakian polytope $P\subset H$. 
In particular notice that the Sasakian K\"ahler class $[\omega_{\mathrm{Sasakian}}]$ satisfying (\ref{fixedclass}) has
\bea\label{lambdaSas}
\lambda_a &= &- \frac{1}{2b_1}~, \qquad a=1,\ldots,d~.
\eea
Changing the transverse K\"ahler class $[\omega_{\mathrm{Sasakian}}]\rightarrow [\omega]\in H^2_B(\mathcal{F}_\xi)$ then 
amounts to moving the edges of the Sasakian polytope $P$.
In principle we could apply the toric geometry formalism of  \cite{guillemin1994} to this case by 
  first projecting the Reeb hyperplane $H$ 
in (\ref{ReebH}) onto $\R^2$. However, as in the discussion of area in (\ref{Atriangles}) and (\ref{areaproj}), 
it is more natural and more invariant to write everything directly in terms of three-dimensional quantities. 
The analogue of (\ref{Delta2d}) is then
\bea\label{generalP}
\mathcal{P} &=& \mathcal{P}(\vec{b};\{\lambda_a\}) \ \equiv \ \{\vec{y}\in H(\vec{b}) \ \mid \ (\vec{y}-\vec{y}_0,\v_a) \geq \lambda_a~, \quad a=1,\ldots,d\}~,
\eea
Here $\vec{y_0}$ is the canonical ``origin'' of the polytope $\mathcal{P}$, given by
\bea\label{originP}
\vec{y}_0 &=& \left(\frac{1}{2b_1},0,0\right) \in H~,
\eea
and $H=H(\vec{b})$ is the Reeb hyperplane (\ref{ReebH}).  

Let us first see that (\ref{generalP}) correctly reduces to the Sasakian polytope (\ref{Ppoly}) in the Sasakian case 
(\ref{lambdaSas}). The edge $E_a$ of the Sasakian polytope $P$ joining the $a$th vertex to the $(a+1)$th vertex is 
\bea\label{edgea}
E_a \ = \ \left\{ (1-t) \frac{\u_a}{2(\u_a,\vec{b})} + t \left.\frac{\u_{a+1}}{2(\u_{a+1},\vec{b})}\  \right| \ t\in [0,1] \right\}~,
\eea
where we have used the vertices of $P$ given in (\ref{yaSasakian}). For $\vec{y}\in E_a$ it is then immediate to verify that
\bea\label{fixy0}
(\vec{y}-\vec{y}_0,\v_a) &=& -(\vec{y}_0,\v_a) \ = \ -\frac{1}{2b_1} \ = \ \lambda_a\mid_{\mathrm{Sasakian}}~.
\eea
Here by construction $(\u_a,\v_a)=0=(\u_{a+1},\v_a)$, since $\v_a$ is normal to the facet in $\mathcal{C}$ generated by the 
two edge vectors $\u_a$, $\u_{a+1}$. In fact the condition (\ref{fixy0}), which applies for all $a=1,\ldots,d$, fixes uniquely  the origin $\vec{y}_0$ given by (\ref{originP}). This shows that the Sasakian polytope is 
\bea
P(\vec{b}) &=& \mathcal{P}\left(\vec{b};\left\{\lambda_a=-\tfrac{1}{2b_1}\right\}\right)~.
\eea

The Sasakian volume $\mathrm{Vol}$ is given by (\ref{SasvolP}), where $\mathrm{vol}(P(\vec{b}))$ is the Euclidean area of the 
Sasakian polytope $P$. Replacing the transverse K\"ahler class $[\omega_{\mathrm{Sasakian}}]\rightarrow [\omega]$ then 
simply replaces the area of $P$ by the area of $\mathcal{P}$ in (\ref{generalP}). It follows that the 
 master volume is
\bea\label{genvol}
\mathcal{V} \ = \ \int_{Y_5}\eta\wedge \frac{\omega^2}{2!} \ = \ \frac{(2\pi)^3}{|\vec{b}|}\vol (\mathcal{P}(\vec{b};\{\lambda_a\}))~.
\eea
We may then compute (\ref{genvol}) explicitly precisely as we did in the previous subsection: by finding the vertices $\vec{y}_a$ 
of $\mathcal{P}$, and then using the area formula for a polytope in terms of its vertices. The vertex $\vec{y}_a$ is the intersection
of edge $a$ with edge $a-1$. Thus it simultaneously solves the three equations
\bea
(\vec{y}_a-\vec{y}_0,\v_a) &=& \lambda_a~, \quad (\vec{y}_a-\vec{y}_0,\v_{a-1}) \ = \  \lambda_{a-1}~, \quad 
(\vec{y}_a-\vec{y}_0,\vec{b}) \ = \ 0~.
\eea 
The solution to this is
\bea
\vec{y}_a &=& \vec{y}_0 + \frac{\lambda_a\, \vec{b}\wedge \v_{a-1}-\lambda_{a-1}\, \vec{b}\wedge \v_a }{(\v_{a-1},\v_a,\vec{b})}~.
\eea
These are the positions of the vertices of $\mathcal{P}$, as a function of the $R$-symmetry vector $\vec{b}$ and K\"ahler class parameters $\{\lambda_a\}$. 
Since $\u_a$ is related to $\v_{a-1}$ and $\v_a$ via (\ref{vtou}), we may also write
\bea
(\v_{a-1},\v_a,\vec{b}) &=& (\u_a,\vec{b})~.
\eea
Using (\ref{Atriangles}) and (\ref{areaproj}), we then have
\bea\label{mastervol}
\mathcal{V} &=& \frac{(2\pi)^3}{|\vec{b}|}\vol (\mathcal{P}(\vec{b};\{\lambda_a\}))\,,\\
&=& \frac{(2\pi)^3}{2(\vec{b},\vec{b})}\sum_{a=1}^d\left(\frac{\lambda_a\, \vec{b}\wedge \v_{a-1}-\lambda_{a-1}\, \vec{b}\wedge \v_a }{(\v_{a-1},\v_a,\vec{b})},\frac{\lambda_{a+1}\, \vec{b}\wedge \v_a-\lambda_{a}\, \vec{b}\wedge \v_{a+1} }{(\v_a,\v_{a+1},\vec{b})},\vec{b}\right)~.\nn
\eea
Again, here we cyclically identify $\lambda_{d+1}\equiv \lambda_1$, $\lambda_0\equiv \lambda_d$. Finally, using a 
 cross product identity we can simplify this expression further. In particular the square norm $(\vec{b},\vec{b})$ then cancels, and we obtain the 
expression
\vskip 0.0cm
\bea
\boxed{~\mathcal{V}(\vec{b};\{\lambda_a\}) \, = \, \frac{(2\pi)^3}{2}\sum_{a=1}^d \lambda_a \frac{\lambda_{a-1}(\v_a,\v_{a+1},\vec{b}) - \lambda_a (\v_{a-1},\v_{a+1},\vec{b})+\lambda_{a+1}(\v_{a-1},\v_a,\vec{b})}{(\v_{a-1},\v_a,\vec{b})(\v_a,\v_{a+1},\vec{b})}~,}\nonumber
\eea
\vskip 0.6cm
\noindent as reported in the introduction in (\ref{volform}).
This is our final expression for the master volume (\ref{Vdef}), as a function of $R$-symmetry vector $\vec{b}$ 
and K\"ahler class parameters $\{\lambda_a\}$.  Notice $\mathcal{V}(\vec{b};\{\lambda_a\})$  is  homogeneous degree $-1$ in the $R$-symmetry vector $\vec{b}$, and quadratic and homogeneous degree 2 in the 
$\{\lambda_a\}$.
We emphasize again that the transverse K\"ahler class $[\omega]$, and hence also volume, 
only depend on $d-2$ of the $d$ variables $\{\lambda_a\}$. Taking (\ref{Vdef}) and setting the $\lambda_a$ all equal as in 
(\ref{lambdaSas}), it is straightforward to recover the original toric formula for the Sasakian volume in 
\cite{Martelli:2005tp}. Notice also that the $\{\lambda_a\}$ are not arbitrary: they must be chosen so that 
the transverse K\"ahler class is strictly positive. The space of such K\"ahler classes is called the \emph{K\"ahler cone} 
$\mathcal{K}$, which in our transverse setting also depends on the $R$-symmetry vector, $\mathcal{K}=\mathcal{K}(\vec{b})$.\footnote{There 
is an unfortunate clash of meanings in the terminology \emph{K\"ahler cone}: there is the definition just introduced, which is a 
space of transverse K\"ahler classes (a real $(d-2)$-dimensional cone), but also the metric cone over a Sasakian manifold (\ref{6cone}) is also 
called a K\"ahler cone (which in this paper has real dimension 6). Hopefully the intended meaning will always be clear from the context.}

We end this section by deriving some formulas which will be useful in the following section. 
Using (\ref{omegalambda}) we can also write the master volume as
\bea
\mathcal{V} &=& (2\pi)^2 \sum_{a,b=1}^d \frac{1}{2!}I_{ab} \lambda_a\lambda_b~,
\eea
where the ``intersection numbers'' $I_{ab}$ are defined as
\bea\label{intersection}
I_{ab} \ \equiv \  \int_{Y_5}\eta\wedge c_a\wedge c_b &=& \frac{1}{(2\pi)^2}\frac{\partial^2 \mathcal{V}}{\partial\lambda_a \partial\lambda_b}~.
\eea
This is in fact independent of $\{\lambda_a\}$, since $\mathcal{V}$ is quadratic in $\{\lambda_a\}$.
Of course, for irrational choices of $\vec{b}$, for which the generic orbits of the $R$-symmetry vector $\xi$ are not closed, the $I_{ab}$ 
will not even be rationally related, let alone integer.
Notice that the Sasakian volume (\ref{Sasvolrho}) may  be expressed as
\bea\label{VolV}
\mathrm{Vol} &=&\frac{1}{8b_1^2}\int_{Y_5}\eta\wedge \rho^2 \ = \  \frac{1}{8b_1^2}\sum_{a,b=1}^d\frac{\partial^2 \mathcal{V}}{\partial\lambda_a \partial\lambda_b} \ = \ \frac{(2\pi)^2}{8b_1^2}\sum_{a,b=1}^d I_{ab}~.
\eea

We may also compute integrals of wedge products of other 
cohomology classes using these formulae. In particular using (\ref{omegalambda}) we have
\bea\label{rhoomegaint}
\int_{Y_5}\eta\wedge \rho\wedge\omega &=&  -(2\pi)^2\sum_{a,b=1}^d I_{ab}\lambda_b\ = \ -\sum_{a=1}^d \frac{\partial \mathcal{V}}{\partial \lambda_a}~.
\eea
Notice this is linear in the $\{\lambda_a\}$. We will also need certain integrals over the toric divisors. 
The images of the toric divisors under the moment map are precisely 
the edges of the  polytope $\mathcal{P}$. On the other hand, the Poincar\'e duals to these
are precisely the $c_a$ introduced in (\ref{omegalambda}). Denoting the corresponding $d$ torus-invariant three-manifolds by $S_a \subset Y_5$, we have
\bea\label{SaV}
\int_{S_a} \eta\wedge\omega &=& \int_{Y_5}\eta\wedge \omega \wedge c_a \ = \  -2\pi\sum_{b=1}^d I_{ab} \lambda_b
\ = \ -\frac{1}{2\pi} \frac{\partial \mathcal{V}}{\partial \lambda_a}~.
\eea

It is interesting to note that we have the following vector identity
\bea\label{Rcharge2}
\sum_{a=1}^d \v_a \int_{S_a}\eta\wedge\omega &=& \frac{\vec{b}}{b_1}\frac{1}{2\pi}\int_{Y_5}\eta\wedge \rho\wedge\omega~,
\eea
which using (\ref{rhoomegaint}), (\ref{SaV}) may equivalently be written as
\bea\label{lineariden}
\sum_{a=1}^d \left(\v_a-\frac{\vec{b}}{b_1}\right)\frac{\partial\mathcal{V}}{\partial\lambda_a} &=& 0~.
\eea
Indeed, this is an identity, holding for all $\vec{b}$ and $\{\lambda_a\}$. 
To see this, first notice that the $i=1$ component is trivial, since $v_a^1=1$ for all $a=1,\ldots,d$. 
Next, since (\ref{lineariden}) is linear in $\lambda_a$, this relation is equivalent to the coefficient of $\lambda_a$ being zero, for each $a=1,\ldots,d$. 
On the other hand, using the explicit form of the master volume (\ref{volform})
and a little rearrangement, the coefficient of $\lambda_a$ being zero in (\ref{lineariden})  is equivalent to 

\begin{align}
& (\v_a,\v_{a+1},\vec{b})\,\v_{a-1} - (\v_{a-1},\v_{a+1},\vec{b})\,\v_{a} +  (\v_{a-1},\v_a,\vec{b})\,\v_{a+1} \nonumber\\[2mm]
& \quad  = \  \frac{\vec{b}}{b_1}\Big[(\v_a,\v_{a+1},\vec{b})   -(\v_{a-1},\v_{a+1},\vec{b})+(\v_{a-1},\v_a,\vec{b})\Big]~.
\end{align}
This identity in turn trivially follows after using the vector quadruple product identity
\bea
 (\v_a,\v_{a+1},\vec{b})\,\v_{a-1} - (\v_{a-1},\v_{a+1},\vec{b})\,\v_{a} +  (\v_{a-1},\v_a,\vec{b})\,\v_{a+1} \ = \  
 (\v_{a-1},\v_a,\v_{a+1})\, \vec{b}~,
\eea
on both sides and recalling that $v_a^1=1$. 
The $i=2,3$ components of (\ref{Rcharge2}) are another manifestation of the fact that there are only 
$d-2$ independent K\"ahler class parameters, parametrized by the $\{\lambda_a\}$. Indeed, $\int_{S_a}\eta\wedge\omega$ 
are proportional to the integral of the transverse K\"ahler class over the  toric divisors. 
Only $d-2$ of these $d$ integrals can be linearly independent, and the $i=2,3$ components of (\ref{Rcharge2}) are the 
two linear relations.

Remarkably, as we shall see in the next section, the above formulae are all that we need to impose the geometric dual of $c$-extremization 
for $Y_7$ which are toric $Y_5$ fibred over a Riemann surface $\Sigma_g$! In particular, everything 
may be computed from the master volume $\mathcal{V}$. 

\section{Fibration over a Riemann surface}\label{sec:fibre}

Having studied the geometry of  $Y_5$, we would now like to fibre this over a Riemann surface $\Sigma_g$ 
to obtain the internal seven-manifold $Y_7$ of a type IIB AdS$_3\times Y_7$ supersymmetric geometry.

\subsection{Fibred geometry}\label{sec:twist}

Topologically, we fibre $Y_5$ over $\Sigma_g$ as follows. The fibres $Y_5$ are toric, admitting an isometric $U(1)^3$ action. 
On the other hand, a $U(1)$ bundle over a Riemann surface $\Sigma_g$ is classified topologically by 
its first Chern number $n\in \Z$. The associated complex line bundle is usually denoted $O(n)_{\Sigma_g}$. 
We may then pick three Chern numbers $\vec{n}=(n_1,n_2,n_3)\in\Z^3$, associated to 
each $U(1)\subset U(1)^3$, giving the direct sum of line bundles $O(\vec{n})_{\Sigma_g}\equiv \oplus_{i=1}^3 O(n_i)_{\Sigma_g}$. 
We then
form the associated bundle 
\bea\label{twistedsister}
Y_7  & \equiv & O(\vec{n})_{\Sigma_g}\times_{U(1)^3} Y_5~.
\eea
The notation here means that we use the $U(1)^3$ transition functions 
of $O(\vec{n})_{\Sigma_g}$ to fibre $Y_5$ over the Riemann surface $\Sigma_g$, using the toric action of $U(1)^3$ on $Y_5$.

In more physical terms, we may introduce three $U(1)$ gauge fields $A_i$ on $\Sigma_g$, $i=1,2,3$, 
with curvatures $F_i=\diff A_i$ satisfying
\bea\label{Fi}
\int_{\Sigma_g} \frac{F_i}{2\pi} &=& n_i\in\Z~.
\eea
The fibration (\ref{twistedsister}) then amounts to the replacement
\bea\label{twistvarphi}
\diff\varphi_i & \rightarrow & \diff\varphi_i + A_i~,
\eea
where recall that $\varphi_i$, $i=1,2,3$, are $(2\pi)$-period coordinates on the torus 
$U(1)^3$.
Before proceeding to analyse the consequences of this, it is important to emphasize that 
all the quantities of interest in section \ref{sec:c} depend only on basic cohomology classes. 
This was also true in section \ref{sec:toric}. In particular this means that 
we may use convenient representatives of certain forms in what follows -- any 
representative will suffice, as long as it has the correct basic cohomology class.

Before the twisting in (\ref{twistedsister}) we may write the one-form $\eta$ on $Y_5$ as
\bea\label{etaY5}
\eta &=& 2\sum_{i=1}^3 \sss_i \diff\varphi_i~.
\eea
Here we have denoted
\bea
\sss_i &\equiv & y_i\mid_{r=1}~,
\eea
which are simply the moment map coordinates $y_i$ restricted to $Y_5$.
In particular, notice (\ref{etaY5}) implies the formula (\ref{yi}) for the moment map coordinates $y_i$, which 
are homogeneous degree two under $r\partial_r$. 
Then the twisting (\ref{twistvarphi}) replaces
\bea\label{etatwisted}
\eta &\rightarrow & \eta_{\mathrm{twisted}} \ \equiv \ 2\sum_{i=1}^3 \sss_i (\diff\varphi_i + A_i)~.
\eea
 Notice that
\bea\label{detatwisted}
\diff\eta_{\mathrm{twisted}} &=& 2\sum_{i=1}^3 \diff \sss_i \wedge (\diff\varphi_i + A_i) + 2\sum_{i=1}^3 \sss_i F_i~,
\eea
where $F_i$ is the curvature two-form on $\Sigma_g$ satisfying (\ref{Fi}). 

Now the holomorphic $(3,0)$-form $\Omega_{(3,0)}$ on the cone $C(Y_5)$ over the fibre
 has an explicit dependence $\ex^{\ii\varphi_1}$, since it has charge 1 under $\partial_{\varphi_1}$ \cite{Martelli:2005tp}.
On the other hand, the holomorphic $(4,0)$-form $\Psi$ on $C(Y_7)$ is constructed by taking the wedge product of the canonical holomorphic $(1,0)$-form 
on $\Sigma_g$ with the $(3,0)$-form $\Omega_{(3,0)}$ on the fibre, twisting the latter using $O(\vec{n})_{\Sigma_g}$. Of course, 
the $(1,0)$-form on $\Sigma_g$ is not globally defined in general (unless the genus $g=1$), being a nowhere zero section of 
$O(2g-2)_{\Sigma_g}$. However, the twisting (\ref{twistedsister}) means that $\ex^{\ii\varphi_1}$ is a nowhere zero section of 
$O(n_1)_{\Sigma_g}$. Neither section exists globally, but the \emph{product} does have a global nowhere zero section, 
and hence gives rise to a global $(4,0)$-form $\Psi$ on $C(Y_7)$, precisely if
\bea\label{n1}
n_1 &=& 2-2g~.
\eea
Thus when $\vec{n}=(2-2g,n_2,n_3)$ the cone $C(Y_7)$ has a global $(4,0)$-form, with 
 $(n_2,n_3)\in \Z^2$ being freely specifiable ``flavour'' twisting parameters.

The actual one-form $\eta$ on $Y_7$ for a supersymmetric geometry will be 
$\eta_{\mathrm{twisted}}$ plus a global basic one-form for the $R$-symmetry foliation 
$\mathcal{F}_\xi$ on $Y_7$. It then suffices to use $\eta_{\mathrm{twisted}}$ in place of $\eta$  to evaluate the various 
integrals that appear in section \ref{sec:c}, which we do in section \ref{sec:formulae} below, since
the global basic one-form will not contribute. Similar remarks apply to the formula
$\diff\eta = \rho/b_1$ in (\ref{etadef}), where recall that for a supersymmetric geometry the holomorphic $(4,0)$-form $\Psi$
on $C(Y_7)$ has charge $b_1=2$, and $\rho$ is the Ricci two-form 
for the transverse K\"ahler metric $J$. In general $\diff\eta_{\mathrm{twisted}}$ will not equal 
$\rho/b_1$ as a differential form, but in cohomology $[\diff\eta_{\mathrm{twisted}}]=[\rho/b_1]\in H^2_B(\mathcal{F}_\xi)$, which 
is sufficient for evaluating the integrals in section~\ref{sec:formulae}.

There is a similar discussion for the K\"ahler form. Since $\partial_{\varphi_i}$ are Killing vectors we have $\mathcal{L}_{\partial_{\varphi_i}}\omega=0$, which implies that $\partial_{\varphi_i}\lrcorner\omega$ is closed. Since manifolds $Y_5$ admitting a toric contact structure have $b_1(Y_5)=0$, it follows that
\bea\label{xidef}
\partial_{\varphi_i}\lrcorner\omega &=& -\diff x_i~, \qquad i=1,2,3~,
\eea
where $x_i$ are global functions on $Y_5$, invariant under the torus action. Similarly to (\ref{detatwisted}) we may then define
\bea\label{omegatwisted}
\omega_{\mathrm{twisted}} & \equiv & \sum_{i=1}^3 \diff x_i \wedge (\diff\varphi_i + A_i) + \sum_{i=1}^3 x_i F_i~,
\eea
which is a closed form, as it should be. Up to an irrelevant exact basic two-form, 
the transverse K\"ahler form on $Y_7$ may then be taken to be 
\bea\label{Jtwisted}
J &=& \omega_{\mathrm{twisted}} + A\, \vol_{\Sigma_g} + \mbox{basic exact}~,
\eea
where we normalize $\int_{\Sigma_g}\vol_{\Sigma_g}=1$, and $A$ 
is effectively a K\"ahler class parameter for the Riemann surface. Notice that the $x_i$ introduced in (\ref{xidef}) are only 
defined up to the addition of constants, leading to a corresponding ambiguity in (\ref{omegatwisted}). However, 
this freedom may then be absorbed into the definition of the  constant parameter $A$ in (\ref{Jtwisted}). We conclude by noting the formulae
\bea\label{moments}
\partial_{\varphi_i}\lrcorner \eta \ = \ 2\sss_i~, \qquad 
\partial_{\varphi_i}\lrcorner \diff\eta \ = \ -2\diff \sss_i~, \qquad \partial_{\varphi_i}\lrcorner\omega &=& -\diff x_i~,
\eea
which will be used repeatedly in the next subsection. 

\subsection{Evaluation of supergravity formulae}\label{sec:formulae}

In this section we would like to evaluate the key off-shell supergravity formulae (\ref{Ssusy}), (\ref{constraint}), (\ref{quantize})  in terms 
of toric data of $Y_5$ and the twisting parameters $\vec{n}$. Using the description of the fibred 
geometry in the previous subsection, we may immediately write down the supersymmetric action (\ref{Ssusy}) as
\bea\label{Ssusyagain}
\Ssusy  &=& \int_{Y_{7}} \eta\wedge \rho \wedge \frac{1}{2!}J^{2}\,,\nn\\
&=& A\int_{Y_5} \eta\wedge \rho\wedge \omega + 2\pi \sum_{i=1}^3 n_i\int_{Y_5} \eta\wedge \omega\wedge \left(b_1 \sss_i\, \omega+ x_i\, \rho\right)~.
\eea
Here we have split the integral over $Y_7$ into an integral over the fibres $Y_5$, and an integral over the Riemann surface base 
$\Sigma_g$. As in the discussion around equation (\ref{Vdef}), in a slight abuse of notation we have
denoted the forms
$\eta$ and  $\rho=b_1\diff\eta$ on $Y_7$  and their restriction to the fibres $Y_5$ by the same symbol, in going from 
the first line to the second line in (\ref{Ssusyagain}). 
To evaluate this we 
 have used the formulae (\ref{Jtwisted}), (\ref{omegatwisted}) for $J$, the formula
(\ref{detatwisted}) to evaluate the $\rho=b_1\diff\eta$ term. The flavour twist parameters $n_i$ arise via
the integrals in (\ref{Fi}). Notice that we may immediately use (\ref{rhoomegaint}) to 
write the first term on the second line of  (\ref{Ssusyagain}) in terms of the master volume $\mathcal{V}$. 

The aim of this subsection is to show that the second term in  (\ref{Ssusyagain}) may similarly be written in 
terms of $\mathcal{V}$, with corresponding formulae also for the constraint (\ref{constraint}) and flux 
quantization condition (\ref{quantize}).  Specifically, we claim that
\bea\label{nice1}
\int_{Y_5} \eta\wedge \omega\wedge \left(b_1 \sss_i\, \omega+ x_i\, \rho\right) &=& -b_1 \frac{\partial\mathcal{V}}{\partial b_i}~.
\eea
We may prove this as follows, generalizing some arguments that first appeared in \cite{Martelli:2006yb}. Recall that the master volume 
$\mathcal{V}$ is by definition
\bea
\mathcal{V} &=& \int_{Y_5} \eta \wedge \frac{1}{2!}\omega^2~.
\eea
As explained in section \ref{sec:toric}, this is a function of the $R$-symmetry vector $\xi=\sum_{i=1}^3 b_i\partial_{\varphi_i}$, 
and on the right hand side of (\ref{nice1}) we are taking the partial derivative of $\mathcal{V}$ with respect to the components $b_i$. 
This may in turn then be computed by determining the first order variations of $\eta$ and $\omega$ under a variation of the 
$R$-symmetry vector $\xi$. 

Focusing first on $\eta$, we thus write
\bea
\xi(t) &=& \xi + t\partial_{\varphi_i}~, \qquad \eta(t)\ = \ \eta + t\nu_i~,
\eea
where $t$ is a (small) parameter, and $\nu_i$ is a one-form for each $i=1,2,3$. 
Since by definition
$\xi(t)\lrcorner \eta(t) = 1$ holds for all $t$, we immediately deduce
\bea
\xi\lrcorner \nu_i &=& -\partial_{\varphi_i}\lrcorner \eta \ = \ -2\sss_i~,
\eea
where in the last equality we have used (\ref{moments}). 
It follows that we may write
\bea\label{nut}
\nu_i &=& - 2\sss_i\eta + \nu_i^T~,
\eea
where by definition $\xi\lrcorner \nu_i^T=0$.
Note that  the one-form $\nu_i$ is precisely the first order variation in $\eta$ induced by varying the $R$-symmetry vector 
$\xi$ in the direction of $\partial_{\varphi_i}$, which of course in turn computes the partial derivative $\partial/\partial b_i$. 

Similarly, we may write
\bea
\omega(t) &=& \omega + t \alpha_i~,
\eea
where $\alpha_i$ is a closed two-form for each $i=1,2,3$.
Since $\omega(t)$ is by definition a transverse K\"ahler form for all $t$, we have 
$\xi(t)\lrcorner\omega(t)=0$, which implies 
\bea
\xi\lrcorner \alpha_i & = & - \partial_{\varphi_i}\lrcorner\omega \ = \ \diff x_i~,
\eea
with the last equality again using (\ref{moments}). 
As in (\ref{nut})
we may then write
\bea
\alpha_i &=& \eta\wedge\diff x_i+\alpha_i^T~,
\eea
where $\xi\lrcorner\alpha_i^T=0$. Since $\alpha_i$ is also closed, it follows that
\bea
\alpha_i^T &=& -x_i\diff\eta + \beta_i^T~,
\eea
where $\beta_i^T$ is a basic closed two-form. Of course, we are always free to 
 shift the transverse K\"ahler class $[\omega] \in H^2_B(\mathcal{F}_\xi)$, which precisely corresponds to the freedom in choosing 
$[\beta_i^T]\in H^2_B(\mathcal{F}_\xi)$ above. Indeed, $\mathcal{V}=V(\xi;[\omega])$ is a function both of $\xi$ and $[\omega]\in H^2_B(\mathcal{F}_\xi)$. 
If we wish to compute the variation induced by varying the $R$-symmetry vector, keeping the $\{\lambda_a\}$ that 
parametrize the transverse K\"ahler class fixed,
 then $\beta_i^T$ is basic exact, {\it i.e.} 
$\beta_i^T=\diff \gamma_i^T$. 

Putting all this together, we compute
\bea
\frac{\partial \mathcal{V}}{\partial b_i} &=& \frac{\diff}{\diff t}\left.\int_{Y_5}\eta(t)\wedge \frac{1}{2!}\omega(t)^2\right|_{t=0}\,,\nn\\
&=& \int_{Y_5}\left( \nu_i \wedge \frac{1}{2!}\omega^2+\eta \wedge \omega\wedge \alpha_i\right) \ = \ \int_{Y_5}\left(-2\sss_i\eta \wedge \frac{1}{2!}\omega^2 - \eta\wedge \omega \wedge x_i\diff\eta\right)\,,\nn\\
&=& -\int_{Y_5} \eta\wedge \omega\wedge \left(\sss_i\, \omega+ \frac{x_i}{b_1}\, \rho\right)~,
\eea
where the third equality uses the transverse form of Stokes' theorem, with $\beta_i^T=\diff\gamma_i^T$ exact and $\omega$ closed, and 
in the very last step we have used $\diff\eta = \rho/b_1$. We have thus proven the desired relation (\ref{nice1}), and conclude that
we may write the supersymmetric action (\ref{Ssusyagain}) as
\bea\label{Ssusynice}
\Ssusy  &=& -A \sum_{a=1}^d \frac{\partial\mathcal{V}}{\partial \lambda_a} - 2\pi b_1 \sum_{i=1}^3 n_i \frac{\partial\mathcal{V}}{\partial b_i}~.
\eea
Recall that a supersymmetric solution necessarily has $b_1=2$, so that the holomorphic $(4,0)$-form $\Psi$ on $C(Y_7)$ has charge $b_1=2$ under the $R$-symmetry 
vector. However, the partial derivative $\partial\mathcal{V}/\partial b_i$ in the second term in (\ref{Ssusynice}) involves regarding 
$\mathcal{V}$ as a function of $\vec{b}=(b_1,b_2,b_3)$, and one only sets $b_1=2$ after taking this derivative.

We next turn to the constraint equation (\ref{constraint}). Evaluating in a similar way to the action (\ref{Ssusyagain}), we compute
\bea\label{constraintagain}
0 &=& \int_{Y_7} \eta\wedge  \rho^2\wedge J\,, \nn\\ 
&  = & A\int_{Y_5}\eta\wedge \rho^2 + 2\pi \sum_{i=1}^3 n_i \int_{Y_5} 
 \eta\wedge \rho \wedge \left(4b_1 \sss_i \omega + x_i\rho\right)~.
\eea
Now recall from (\ref{rhoomegaint}) that 
\bea\label{rhoeomegaintagain}
\int_{Y_5}\eta\wedge \rho\wedge\omega &=& -\sum_{a=1}^d \frac{\partial \mathcal{V}}{\partial \lambda_a}~.
\eea
As in the computation of the action above, we then take the 
 partial derivative with respect to the $R$-symmetry vector
\bea
\frac{\partial}{\partial b_i}\sum_{a=1}^d \frac{\partial \mathcal{V}}{\partial \lambda_a} &=& -\frac{\partial}{\partial b_i}\int_{Y_5}\eta\wedge b_1\diff\eta \wedge \omega~,
\eea
where we have replaced $\rho=b_1\diff\eta$. There is hence an explicit $b_1$-dependence in the integrand, which leads to
\bea
\frac{\partial}{\partial b_i}\sum_{a=1}^d \frac{\partial \mathcal{V}}{\partial \lambda_a} \ =\  \int_{Y_5}\left( 4 b_1 \sss_i \eta\wedge \diff\eta \wedge \omega +\eta\wedge \diff\eta \wedge x_i\rho\right)  
- \delta_{1,i} \int_{Y_5}\eta\wedge\diff\eta\wedge\omega~.
\eea
Using also (\ref{VolV}) for the first term in the constraint equation (\ref{constraintagain}), the latter thus reads 
\bea\label{constraintnice}
0 &=& A \sum_{a,b=1}^d \frac{\partial^2\mathcal{V}}{\partial\lambda_a\partial\lambda_b} - 2\pi n_1 \sum_{a=1}^d\frac{\partial \mathcal{V}}{\partial\lambda_a} + 2\pi b_1 \sum_{a=1}^d \sum_{i=1}^3 n_i\frac{\partial^2 \mathcal{V}}{\partial \lambda_a\partial b_i}~,
\eea
where we have again used (\ref{rhoeomegaintagain}).

Finally, we turn to the flux quantization condition (\ref{quantize}). Recall here that $\Sigma_A\subset Y_7$ form a representative basis of five-cycles, forming a basis for the free part of $H_5(Y_7;\Z)$, 
where the representative submanifolds should be tangent to the $R$-symmetry vector $\xi$. A distinguished such five-cycle is a copy of the fibre $Y_5$ at a fixed point on the Riemann surface 
base $\Sigma_g$. Denoting the corresponding five-form flux quantum number by $N\in\mathbb{Z}$, in this case (\ref{quantize}) reads
\bea\label{Nnice}
\frac{2(2\pi \ell_s)^4g_s}{L^4}N &=& \int_{Y_5}\eta\wedge\rho\wedge\omega \ = \ -\sum_{a=1}^d\frac{\partial\mathcal{V}}{\partial\lambda_a}~.
\eea
Being the five-form flux through $Y_5$, it is natural to interpret $N$ as the number of D3-branes that we are wrapping on the Riemann surface 
$\Sigma_g$.\footnote{Although $N$ is an integer, it is not necessarily an \emph{arbitrary} integer, as we shall see in examples later in the paper. 
Ultimately this is related to the fact that the fibre class $[Y_5]\in H_5(Y_7;\Z)$ can be non-primitive, {\it i.e.} there can exist an integer 
$n>1$ such that $[Y_5]/n \in H_5(Y_7;\Z)$ is still an integer class. In this case $N$ is necessarily divisible by $n$.} The remaining five-cycles $\Sigma_a$ in $Y_7$ are generated by torus-invariant three-manifolds $S_a\subset Y_5$ fibred over $\Sigma_g$, 
where $a=1,\ldots,d$. Denoting the corresponding five-form flux quantum numbers by $M_a\in\mathbb{Z}$, the quantization condition 
 (\ref{quantize}) reads
\bea\label{Maint}
\frac{2(2\pi \ell_s)^4g_s}{L^4}M_a &=&\int_{\Sigma_a} \eta\wedge \rho\wedge J\,,  \nn\\ &=& A \int_{S_a} \eta\wedge \rho + 2\pi \sum_{i=1}^3 n_i \int_{S_a}\eta\wedge  (2b_1\sss_i \omega + x_i\rho)~.
\eea
Using (\ref{SaV}) we then compute
\bea
\frac{\partial}{\partial b_i}\frac{\partial\mathcal{V}}{\partial\lambda_a} &=& -2\pi \frac{\partial}{\partial b_i}\int_{S_a}\eta\wedge\omega\ = \  2\pi\int_{S_a}\eta\wedge \left(2\sss_i \omega +\frac{x_i}{b_1}\rho\right)~.
\eea
Thus the flux quantization condition (\ref{Maint}) reads
\bea\label{Manice}
\frac{2(2\pi \ell_s)^4g_s}{L^4}M_a &=& \frac{A}{2\pi} \sum_{b=1}^d \frac{\partial^2\mathcal{V}}{\partial\lambda_a\partial\lambda_b} + b_1 \sum_{i=1}^3 n_i \frac{\partial^2 \mathcal{V}}{\partial\lambda_a\partial b_i}~.
\eea

The toric three-cycles $[S_a]\in H_3(Y_5,\Z)$ are not independent in $H_3(Y_5,\Z)$. Indeed, there are $d$ toric three-cycles, $a=1,\ldots,d$, but 
$\dim H_3(Y_5,\R) = d-3$. Although $\{[S_a]\}$ generate the free part of $H_3(Y_5,\Z)$, they must then necessarily satisfy
$3$ relations. These are \cite{Franco:2005sm}
\bea\label{homSa}
\sum_{a=1}^d v_a^i [S_a] &=& 0\in H_3(Y_5;\Z)~, \qquad i=1,2,3~,
\eea
where $\{\v_a\}$ are the inward pointing normals to the facets of the moment map polyhedral cone. 
Fibering each $S_a$ over $\Sigma_g$ gives rise to a torus-invariant five-manifold $\Sigma_a\subset Y_7$. We then 
have the corresponding homology relation for five-cycles in $Y_7$:
\bea\label{homrel}
\sum_{a=1}^d v_a^i [\Sigma_a] &=& -n_i [Y_5]\in H_5(Y_7;\Z)~, \qquad i=1,2,3~.
\eea
In particular this immediately implies that
\bea\label{Marel}
\sum_{a=1}^d v_a^i M_a &=& -n_i N~,\qquad i=1,2,3~,
\eea
leading in general to $d-2$ independent flux quantum numbers (the $d-3$ independent $\{M_a\}$, together with $N$). 

Rather than prove (\ref{homrel}), we instead present an elementary derivation of 
 (\ref{Marel}), which is all that we will need for the present paper. Starting with the identity (\ref{lineariden}), which recall holds for all $\vec{b}$ and all $\{\lambda_a\}$,
taking derivatives leads to
\bea\label{secondiden}
\sum_{a,b=1}^d v_a^i  \frac{\partial  \mathcal{V}}{\partial \lambda_a \partial \lambda_b } & = &    \frac{b_i}{b_1} \sum_{a,b=1}^d  \frac{\partial^2   \mathcal{V}}{\partial \lambda_a \partial \lambda_b } ~,\nn\\
 \qquad \sum_{a=1}^d v_a^i  \frac{\partial^2  \mathcal{V}}{\partial b_j \partial \lambda_a } & = & \left(   \frac{\delta_{ij}}{b_1}    - \frac{b_i \delta_{1j}}{b_1^2} \right)\sum_{a=1}^d  \frac{\partial  \mathcal{V}}{\partial \lambda_a }  
+ \frac{b_i}{b_1} \sum_{a=1}^d   \frac{\partial^2  \mathcal{V}}{\partial b_j \partial \lambda_a }~.
\eea
Multiplying (\ref{Manice}) by $v_a^i$ and summing over $a=1,\ldots,d$ then gives
\bea
\frac{2(2\pi \ell_s)^4 g_s}{L^4} \sum_{a=1}^d v_a^iM_a &=&   \frac{A}{2\pi}\sum_{a,b=1}^d v_a^i \frac{\partial^2\mathcal{V}}{\partial\lambda_a\partial\lambda_b}+b_1\sum_{a=1}^d \sum_{j=1}^3 v_a^i n_j\frac{\partial^2 \mathcal{V}}{\partial\lambda_a\partial b_j}~.
\eea
The right hand side may be evaluated using (\ref{secondiden}). The term proportional to $A$ may be eliminated using 
the constraint equation (\ref{constraintnice}) to give
\bea
&&\frac{2(2\pi \ell_s)^4 g_s}{L^4} \sum_{a=1}^d v_a^iM_a \ =\   \frac{b_i}{b_1}n_1 \sum_{a=1}^d\frac{\partial \mathcal{V}}{\partial\lambda_a} -  b_i \sum_{a=1}^d \sum_{j=1}^3 n_j\frac{\partial^2 \mathcal{V}}{\partial \lambda_a\partial b_j}\nn\\
&& \ \qquad \ \ + \left(n_i-\frac{b_i n_1}{b_1}\right)\sum_{a=1}^d  \frac{\partial  \mathcal{V}}{\partial \lambda_a } +{b_i}\sum_{a=1}^d 
\sum_{j=1}^3 n_j  \frac{\partial^2  \mathcal{V}}{\partial b_j \partial \lambda_a }\ = \ n_i \sum_{a=1}^d  \frac{\partial  \mathcal{V}}{\partial \lambda_a } ~.
\eea
Using the relation (\ref{Nnice}), one 
finally deduces (\ref{Marel}).

\subsection{Summary}\label{sec:summary}

We may now summarize the procedure for carrying out the geometric dual of $c$-extremization in \cite{Couzens:2018wnk}, 
for toric $Y_5$ fibred over a Riemann surface $\Sigma_g$.

First choose a Gorenstein toric K\"ahler cone $C(Y_5)$, with toric data $\{\v_a=(1,\w_a)\in\Z^3\mid a=1,\ldots,d\}$. 
We may then compute the master volume function
\bea
\label{mvolfn}
\mathcal{V}(\vec{b};\{\lambda_a\})   =  \frac{(2\pi)^3}{2}\sum_{a=1}^d \lambda_a \frac{\lambda_{a-1}(\v_a,\v_{a+1},\vec{b}) \! -\! \lambda_a (\v_{a-1},\v_{a+1},\vec{b})\!+\!\lambda_{a+1}(\v_{a-1},\v_a,\vec{b})}{(\v_{a-1},\v_a,\vec{b})(\v_a,\v_{a+1},\vec{b})}~.
\eea
This is a function of both the trial $R$-symmetry vector $\vec{b}=(b_1,b_2,b_3)$, and also the transverse K\"ahler class 
parameters $\{\lambda_a\mid a=1,\ldots,d\}$. Two of the latter $d$ variables are redundant, so that in general $\mathcal{V}$ 
is a function of $3+d-2=d+1$ parameters.

The fibration over $\Sigma_g$ is specified by the flavour twisting parameters $\vec{n}=(2-2g,n_2,n_3)$.
We next impose the constraint equation
\bea\label{constraintformula} 
0 &=& A \sum_{a,b=1}^d \frac{\partial^2\mathcal{V}}{\partial\lambda_a\partial\lambda_b} - 2\pi n_1 \sum_{a=1}^d\frac{\partial \mathcal{V}}{\partial\lambda_a} + 2\pi b_1 \sum_{a=1}^d \sum_{i=1}^3 n_i\frac{\partial^2 \mathcal{V}}{\partial \lambda_a\partial b_i}~,
\eea
and flux quantization conditions
\bea\label{Nformula}
\frac{2(2\pi \ell_s)^4 g_s}{L^4}N &=& -\sum_{a=1}^d \frac{\partial \mathcal{V}}{\partial\lambda_a}~,\\
\label{Maformula}
\frac{2(2\pi \ell_s)^4 g_s}{L^4}M_a &=& \frac{A}{2\pi}\sum_{b=1}^d\frac{\partial^2\mathcal{V}}{\partial\lambda_a\partial\lambda_b}+b_1\sum_{i=1}^3 n_i\frac{\partial^2 \mathcal{V}}{\partial\lambda_a\partial b_i} ~.
\eea
Here $A$ is effectively an additional K\"ahler class parameter for $\Sigma_g$, making $d+2$ parameters in total. As explained at the end of the previous subsection, the $\{M_a\mid a=1,\ldots,d\}$ comprise 
$d-3$ independent flux quantum numbers, due to the topological relation (\ref{Marel}). Thus generically equations
(\ref{constraintformula}), (\ref{Nformula}), (\ref{Maformula}) impose $1+1+d-3=d-1$ relations. 

Finally, we set $b_1=2$ so that the holomorphic $(4,0)$-form $\Psi$ has charge $b_1=2$ under the $R$-symmetry vector.
In total we have then imposed $d$ relations on the $d+2$ parameters, so that the resulting action
\bea\label{Ssusyformula}
\Ssusy  &=& -A \sum_{a=1}^d \frac{\partial\mathcal{V}}{\partial \lambda_a} - 2\pi b_1 \sum_{i=1}^3 n_i \frac{\partial\mathcal{V}}{\partial b_i}~,
\eea
is in general a function of two remaining variables.\footnote{An exception to this counting 
is the untwisted $g=1$ examples studied in \cite{Couzens:2018wnk}. We discuss this case further in section \ref{secypqdp2}.}
Extremizing this action or, equivalently, the off-shell central charge 
\bea\label{cS24}
\cZ & = & \frac{3L^8}{(2\pi)^6g_s^2\ell_s^8} \Ssusy~,
\eea
we then have that its value at the critical point gives the on-shell central charge
\bea\label{cZonshell}
\csugra & = &\left. \cZ\right|_{\mathrm{on-shell}}\,.
\eea

As in \cite{Couzens:2017nnr, Couzens:2018wnk} we may also compute the $R$-charges $R_a=R[S_a]$ of baryonic operators dual to D3-branes wrapping the supersymmetric three-manifolds
$S_a\subset Y_7$, at a fixed point on the base $\Sigma_g$. These are 
given\footnote{The expression for the $R$-charges in \cite{Couzens:2017nnr}
were obtained using a calibration type argument. The consistency of the results in 
\cite{Couzens:2018wnk} and in this paper provide overwhelming
evidence for its veracity. Nevertheless, it would be desirable to provide a direct proof
using $\kappa$-symmetry.} by the general formula \cite{Couzens:2017nnr}
\bea\label{rchgegenexpgeneral}
R_a\ = \ R[S_a] &= &
\frac{L^4}{(2\pi)^3\ell_s^4g_s}\int_{S_a}\eta\wedge \omega \ = \ -\frac{L^4}{(2\pi\ell_s)^4g_s}\frac{\partial\mathcal{V}}{\partial\lambda_a}~,
\eea
where we have used (\ref{SaV}) in the second equality. Note that (\ref{Nformula}) then implies that
\bea\label{R2N}
\sum_{a=1}^d R_a &=& 2N~.
\eea
For the quiver gauge theories we discuss in examples later in the paper, this last relation has the simple interpretation 
that each term in the superpotential has $R$-charge 2. The same relation holds for the parent AdS$_5\times Y_5$ 
solutions \cite{Franco:2005sm}. In fact using (\ref{Rcharge2}) equation (\ref{R2N}) is simply the $i=1$ component 
of the relation
\bea
\sum_{a=1}^d \v_a R_a &=& 2\frac{\vec{b}}{b_1}N \ = \ \vec{b} N~,
\label{summarsv}
\eea
in the last step setting $b_1=2$. Compare this to the homology relations (\ref{Marel}). The relation 
(\ref{summarsv}) will be of practical use later, in relating $\cZ$-extremization 
in gravity to $c$-extremization in the dual field theory.

\subsection{Existence of solutions}\label{sec:exist}

The $\cZ$-extremization procedure just summarized determines the central charge (\ref{cZonshell}) and $R$-charges of BPS baryonic operators (\ref{rchgegenexpgeneral})
in gravity, \emph{assuming} such a solution exists. In this section we elaborate further on this point.

It is instructive to first compare our $\cZ$-extremization problem to volume minimization \cite{Martelli:2005tp, Martelli:2006yb}. 
As we recalled after equation (\ref{Sasvolrho}), here the 
 Reeb vector $\vec{b}$ for a Sasaki-Einstein metric extremizes the Sasakian volume $\mathrm{Vol}=\mathrm{Vol}(\vec{b})$, subject 
to the constraint $b_1=3$. The Sasakian volume is easily shown to be strictly convex, and tends to $+\infty$ as one 
approaches the boundary of the Reeb cone $\partial \mathcal{C}^*$ from the interior. From this one can prove there always exists a unique 
critical point $\vec{b}\in \mathcal{C}^*_{\mathrm{int}}$, which minimizes the volume. It follows that volume minimization 
determines the unique Reeb vector and Sasaki-Einstein volume, \emph{assuming} such a Sasaki-Einstein metric exists. The latter is then a 
problem in PDEs. In this toric geometry setting, it was later proven in \cite{Futaki:2006cc} that the relevant PDE always admits a solution, 
thus fully solving the toric Sasaki-Einstein problem. The non-toric case is more involved: there can be obstructions 
to the existence of a Sasaki-Einstein metric for a Reeb vector that minimizes the volume; for example, those discussed in \cite{Gauntlett:2006vf}. In this 
case the minimized volume is \emph{not} the volume of a Sasaki-Einstein manifold, since the latter doesn't exist! 
In fact much more can now be said about this general existence problem \cite{Collins:2015qsb}.

Similar issues arise for $\cZ$-extremization, although the situation is more involved. Recall that our construction in section \ref{sec:toric}
required the $R$-symmetry vector to be inside the Reeb cone, $\vec{b}\in\mathcal{C}^*_{\mathrm{int}}$, and the transverse K\"ahler 
class $[\omega]$ determined by the $\{\lambda_a\}$ to be inside the K\"ahler cone $\mathcal{K}(\vec{b})$. If either of these 
don't hold, the polytope $\mathcal{P}$ in (\ref{generalP}) is not well-defined. However, the formulae in 
section \ref{sec:summary}, including the master volume (\ref{mvolfn}), make sense for generic $\vec{b}$, $\{\lambda_a\}$. 
After carrying out the extremal problem, one then has to {\it a posteriori} check that the critical 
values of $\vec{b}$ and the transverse K\"ahler class determined by $\{\lambda_a\}$ indeed lie inside their respective Reeb and K\"ahler cones. 
It is clear already from the examples studied in \cite{Couzens:2018wnk} that this is not necessarily the case,  the conclusion then being 
that such supergravity solutions do not exist. For example, a simple diagnostic is to look at the final central charge (\ref{cZonshell}) 
and $R$-charges (\ref{rchgegenexpgeneral}). If the critical $R$-symmetry vector and transverse K\"ahler class lie inside the Reeb and K\"ahler cone, respectively, 
these quantities are guaranteed to all be positive. Thus if one is negative, the solution cannot exist. Of course, this also has a straightforward 
dual interpretation in field theory, where the central charge and $R$-charges of BPS operators should all be positive. 
This situation should however be contrasted with volume minimization for Sasaki-Einstein metrics, where there always exists a unique critical Reeb vector inside the Reeb cone, 
with then a necessarily positive critical volume.

In practice one thus needs to check some positivity conditions after performing $\cZ$-extremization. We leave a general analysis of this problem 
for the future. However, one still needs to show existence of a solution to the PDE (\ref{boxReqn}), analogous to the Einstein equation in the Sasaki-Einstein setting. 
Being at an extremum of $\cZ$ can be viewed as a necessary global condition for this PDE to admit a solution, but more generally one wants to know if this is also sufficient. 
Given the situation in Sasaki-Einstein geometry, it is natural to conjecture that there are no further obstructions to solving this PDE in the toric case, but 
more generally one might expect more exotic obstructions, with some final picture close to the K-stability of   \cite{Collins:2015qsb} for Sasaki-Einstein manifolds. 
Finally, as discussed in \cite{Couzens:2018wnk}, it is clear that for general supergravity solutions, starting with a \emph{convex} polyhedral cone $\mathcal{C}$ 
is too strong. We will  recall this in relation to certain examples in section \ref{secypqexamples}, and again at the beginning of section 
\ref{secexplexamples}.
However, such geometries no longer 
have any obvious relation to fibering Sasaki-Einstein geometries over a Riemann surface, and hence to dual D3-brane quiver gauge theories wrapped 
on that Riemann surface. Such supergravity solutions certainly exist, but there is currently no conjecture for the dual $(0,2)$ SCFT.

Most of these questions are clearly well beyond the scope of this paper. In the remainder of the paper we apply the formalism summarized in section \ref{sec:summary} to a variety of examples, recovering 
results for various explicit supergravity solutions, 
and comparing to $c$-extremization in the field theory duals. Remarkably, we will see in examples, and conjecture more generally, that 
the off-shell $\cZ$-function is directly related to the off-shell trial $c$-function in field theory, thus leading to a (formal) matching
between $\cZ$-extremization and $c$-extremization.
The positivity and existence questions raised in this subsection are then reflected in the dual field theory as whether or not the
putative  IR superconformal fixed point actually exists.

\section{The universal twist revisited}
\label{unitwist}

As a warm-up we will begin by applying our general formalism to the case often referred to as the \emph{universal twist} in the literature. Specifically, we consider a seven-dimensional manifold $Y_7$ that is a fibration of a toric $Y_5$ over a genus $g>1$ 
Riemann surface $\Sigma_g$, where the twisting is only along the $U(1)_R$ $R$-symmetry. The corresponding supergravity solutions exist for any $Y_5=SE_5$ that is a quasi-regular Sasaki-Einstein manifold, 
and were constructed  in \cite{Gauntlett:2006qw}. Here the six-dimensional transverse K\"ahler metric in (\ref{metric}) is 
simply a product $H^2/\Gamma\times KE_4$, where $H^2/\Gamma$ is a constant negative curvature 
Riemann surface $\Sigma_{g>1}$, and $KE_4$ denotes the positive curvature K\"ahler-Einstein orbifold  $SE_5/U(1)_R$. In particular this product 
of Einstein metrics solves the PDE (\ref{boxReqn}) in a trivial way, where the total Ricci scalar $R$ 
equals a positive constant.
The gravitational central charges for these solutions, calculated in \cite{Gauntlett:2006qw}, were shown to agree precisely with the central charges obtained from $c$-extremization in the dual two-dimensional $(0,2)$ field theories in \cite{Benini:2015bwz}. Below we will show that indeed our formulas reduce to combinations of the corresponding toric Sasakian formulas, thus making direct contact with the results of \cite{Martelli:2005tp}. We will also compute the $R$-charges of the 
toric three-cycles $S_a$,   highlighting the fact that they need to obey a simple quantization condition. 

Let us start by recalling the expressions for the toric Sasakian volume and the volumes of the toric three-cycles. These are given by  \cite{Martelli:2005tp}
\bea
\mathrm{Vol} (Y_5) & =&  \frac{\pi^3}{b_1}  \sum_a \frac{(\v_{a-1},\v_a,\v_{a+1})}{(\v_{a-1},\v_a,\vec{b})(\v_a,\v_{a+1},\vec{b})} ~,
\label{sasakY}\\
\mathrm{Vol} (S_a) &=&  2\pi^2  \frac{(\v_{a-1},\v_a,\v_{a+1})}{(\v_{a-1},\v_a,\vec{b})(\v_a,\v_{a+1},\vec{b})} ~,
\label{sasaksigma}
\eea
respectively. In terms of the master volume $\mathcal{V}$, we have 
\bea\label{toricsavol}
\sum_{a,b=1}^d\frac{\partial^2 \mathcal{V}}{\partial\lambda_a \partial\lambda_b}  & = & 8 b_1^2  \mathrm{Vol} (Y_5)~, \qquad 
\sum_{b=1}^d \frac{\partial^2   \mathcal{V}}{\partial \lambda_a \partial \lambda_b} \ = \ 4\pi b_1 \mathrm{Vol} (S_a) ~,
\eea
where the first equation has already appeared in (\ref{VolV}).
The universal twist corresponds to choosing the fluxes $n_i$ to be
aligned with the $R$-symmetry vector, namely we require
\bea\label{twistb}
n_i  & =  &   \frac{n_1}{b_1} b_i~,
\eea
with $n_1=2-2g$ as in \eqref{n1}.
Note that we will need to check, {\it a posteriori}, that after carrying out $c$-extremization
the on-shell value of $\vec{b}$ is consistent with the left hand side of (\ref{twistb})  being integers. 
Inserting this into the formulas for the action (\ref{Ssusyformula}),  the constraint (\ref{constraintformula})    and the flux quantization condition (\ref{Maformula}), and using the fact that the master volume $\mathcal{V}$ is homogeneous of degree minus one in $\vec{b}$,
these reduce respectively to 
\bea
\Ssusy &=& A  \frac{2(2\pi \ell_s)^4 g_s}{L^4}  N   + 2\pi n_1   \mathcal{V} ~,\\
\label{constraintformulauniverse}
0 &=& A\sum_{a,b=1}^d \frac{\partial^2\mathcal{V}}{\partial{\lambda_a}\partial{\lambda_b}}+ 4\pi n_1
   \frac{2(2\pi \ell_s)^4 g_s}{L^4}  N~,\\
\label{formuletta}
2  \frac{(2\pi \ell_s)^4 g_s}{L^4}  M_a &=&    \frac{A}{2\pi}\sum_{b=1}^d\frac{\partial^2\mathcal{V}}{\partial\lambda_a\partial\lambda_b}-2n_1 \frac{\partial \mathcal{V}}{\partial\lambda_a}~,
\eea
with
\begin{align}\label{enntwo}
\frac{2(2\pi \ell_s)^4 g_s}{L^4}N\  =\  -\sum_{a=1}^d \frac{\partial \mathcal{V}}{\partial\lambda_a}~.
\end{align}
We can use (\ref{constraintformulauniverse}) to eliminate $A$ from the action, which can then be written as 
\bea
\Ssusy &=& 2\pi n_1  \mathcal{V} -  \left( \frac{(2\pi \ell_s)^4 g_s}{L^4}\right)^2 \frac{2\pi n_1   N^2 }{b_1^2 \mathrm{Vol} (Y_5)} ~.
\eea
Notice that the second term depends only on the $R$-symmetry vector $\vec{b}$ through the inverse Sasakian volume, similarly to the trial central charge of the related 
 four-dimensional problem. On the other hand, the first term is still the general master volume, and thus it depends also on the K\"ahler parameters $\lambda_a$.

Let us now consider the $d$ relations (\ref{formuletta}). Using (\ref{toricsavol}) and the explicit expression for  $\partial \mathcal{V}/{\partial\lambda_a}$, 
we can write this as a linear system
\bea
\label{linearsys}
(2\pi)^2n_1 \sum_{b=1}^d I_{ab}\lambda_b &=& A b_1 \mathrm{Vol} (S_a)-\frac{(2\pi \ell_s)^4 g_s}{L^4}M_a~.
\eea
Here $I_{ab}$ is the intersection matrix (\ref{intersection}) which has rank $d-2$, corresponding to the redundancy of two of the K\"ahler parameters. There is thus no unique solution for the $\lambda_a$ in (\ref{linearsys}). However, one can show that there exists a ``gauge'' in which all the $\lambda_a$ are equal and solve   (\ref{linearsys}).  
In particular, setting $\lambda_a=\lambda$ for $a=1,\dots,d$ we have 
\bea
\mathcal{V}  & = & 4 b_1^2 \lambda^2  \mathrm{Vol} (Y_5)\,,
 \label{putin}
\eea
with the value of $\lambda$ being determined, from \eqref{enntwo}, by the quantization condition 
\bea
\lambda & = & -\frac{(2\pi \ell_s)^4 g_s N}{4L^4 b_1^2  \mathrm{Vol} (Y_5)}~.
\label{trump}
\eea
Inserting this into the action, the latter then reads
\bea
\Ssusy  & = & -  \left( \frac{(2\pi \ell_s)^4 g_s}{L^4}  \right)^2  \frac{3n_1\pi  N^2}{ 2b_1^2 \mathrm{Vol} (Y_5)}~.
\eea

As in the Sasakian setting, this action has to be extremized with respect to $b_2, b_3$, holding $b_1$ fixed. However, presently we have to set $b_1=2$, while in the Sasaki-Einstein case we have $b_1=3$. Defining 
$\vec{b} = \frac{2}{3}\vec{r}$ and using the fact that $\Vol (Y_5)$ is homogeneous of degree minus three in $\vec{b}$, we can rewrite the action as
\bea
\Ssusy  (\vec{r}) & = & -  \left( \frac{(2\pi \ell_s)^4 g_s}{L^4}  \right)^2    \frac{\pi n_1 N^2}{(\tfrac{3}{2})^2 b_1^2\mathrm{Vol} (Y_5) (\vec{r})}~.
\eea
Since $\mathrm{Vol} (Y_5) (\vec{r})$ with $r_1=3$ is extremized by the critical Reeb vector $\vec{r}=\vec{r}_*$, with $\mathrm{Vol} (Y_5) (\vec{r}_*)$ being the Sasaki-Einstein volume, we conclude that $\Ssusy  (\vec{r})$ is extremized 
for the critical $R$-symmetry vector given by
\begin{align}\label{critbv}
\vec{b}_* \ = \ \frac{2}{3}\vec{r}_*\,.
\end{align}
The value of the trial central charge  at the critical point is then
\bea
\cZ  |_\mathrm{on-shell}
 & =  & - \frac{4n_1\pi^3  N^2}{3\mathrm{Vol} (Y_5) (\vec{r}_*)}~,
\eea
where the last step uses (\ref{cS}).
Finally, recalling the standard relation between the $a$ central charge of the four-dimensional SCFT and the volume of the Sasaki-Einstein manifold of the corresponding AdS$_5\times Y_5$ type IIB solution,
\bea
a^{4d} & = & \frac{\pi^3N^2}{4 \mathrm{Vol} (Y_5) (\vec{r}_*)}~,
\eea
we  obtain the relation
\bea
\csugra \ = \ \cZ  |_\mathrm{on-shell} & =  &     \frac{32}{3}(g-1)a^{4d} ~,
\eea
in agreement with the explicit supergravity solutions \cite{Gauntlett:2006qw}
and $c$-extremization in the two-dimensional $(2,0)$ SCFTs \cite{Benini:2015bwz}.

It is straightforward to compute the  geometric $R$-charges, which read
\bea
R_a & = &         \frac{\pi N   \mathrm{Vol} (S_a) (\vec{r}_*) }{3  \mathrm{Vol} (Y_5) (\vec{r}_*)} \ = \ N R_a^{4d}~,
\eea
where $R_a^{4d}$ denote the (geometric) $R$-charges of the four-dimensional theories (which
are usually defined without the factor of $N$). 
This is in agreement with the field theory results and with  the explicit  gravitational solutions   
\cite{Couzens:2017nnr}. Note that  (\ref{formuletta}) relates  the $R$-charges  to the integer fluxes $M_a$,  as 
\bea\label{MaRa}
M_a & = &  (g-1) N  R_a^{4d}~,
\eea
implying in particular that the $R$-charges of the  parent four-dimensional theory must be  rational numbers. 
This is a manifestation of the fact that, as discussed in \cite{Gauntlett:2006qw}, 
the Sasaki-Einstein manifolds must be quasi-regular
with Reeb vector $\vec{r}_*\in\mathbb{Q}^3$ so that the exact $R$-symmetry $\vec{b}_*$
of the four-dimensional theory in \eqref{critbv} generates a $U(1)$ action on $Y_5$ and the twisting (\ref{twistb}) is well-defined.\footnote{Note that if for a given Sasaki-Einstein manifold  and genus $g$ the $n_i$ in (\ref{twistb}) are not integers, then we can consider taking an orbifold of the Sasaki-Einstein
space, as discussed in \cite{Gauntlett:2006qw}.} 
Conversely, for a fixed quasi-regular Sasaki-Einstein manifold with spectrum of $R$-charges $\{R_a^{4d}\in\mathbb{Q}\}$, 
since the left hand side of (\ref{MaRa}) must be integers, this leads to a corresponding divisibility condition on the integer $(g-1)N$. 


\section{$SE_5$ quiver theories reduced on $\Sigma_g$}
\label{secypqdp2}

It is straightforward to apply our general formalism to examples of $Y_7$ arising as a fibration
of $Y_5$ over $\Sigma_g$, with $Y_5$ given by toric Sasaki-Einstein spaces, $SE_5$, which have more general twisting
than the universal twist considered in the last section. To illustrate we will consider
$Y_5=Y^{p,q}$, and $Y_5=X^{p,q}$. 
These include $Y^{2,1}$ and $X^{2,1}$, which are the $SE_5$ manifolds associated with the canonical complex 
cones over the first and second del Pezzo surfaces, $dP_1$, $dP_2$, respectively. 
In each case we calculate the off-shell trial central charge, $\cZ$, and 
then extremize to obtain the on-shell central charge, $\csugra=\cZ  |_\mathrm{on-shell}$, and $R$-charges. Furthermore,
we also show how these results explicitly agree with 
$c$-extremization in the dual $d=4$ quiver gauge theories, dual to
the $SE_5$, after reducing on $\Sigma_g$ with suitable twist. In fact, we will see that, generically, there is
actually an off-shell agreement between the field theory trial central charge after extremizing over
the baryon mixing, and the trial central charge arising in the geometry computation. A similar off-shell agreement also arises for the $R$-charges.
As discussed in section \ref{sec:exist},  the matching between our geometric results and field theory is {\it a priori} a formal
matching, since it assumes existence of the supergravity solution. We discuss some of these issues further also at 
 the beginning of section \ref{secexplexamples}.

\subsection{$Y^{p,q}$}\label{secypqexamples}
The $Y^{p,q}$ Sasaki-Einstein metrics were first constructed in \cite{Gauntlett:2004yd},  and the associated toric data was derived in \cite{Martelli:2004wu}. 
The inward pointing normal vectors are given by\bea\label{Ypqv}
\v_1 \,  =\,  (1,0,0)~, \quad \v_2 \, = \, (1,1,0)~, \quad \v_3 \, = \, (1,p,p)~, \quad \v_4 \, = \, (1,p-q-1,p-q)~.
\eea
Notice that these are properly ordered in an anti-clockwise direction. Furthermore, the 
Sasaki-Einstein metrics have $p>q>0$ and the polyhedral cone with vectors $\v_a$ is convex.
We want to consider $Y_7$ which are obtained by a fibration of $Y^{p,q}$ over $\Sigma_g$.
As usual, we take $n_1=2(1-g)$, and for simplicity we restrict to considering flavour fluxes that preserve the $SU(2)$ symmetry and set\footnote{Note that we should not set $b_2=b_3$ at this stage, but we will derive this condition from extremization. } $n_2=n_3$.
It will be convenient to separate the cases when the genus $g\neq 1$  from the case when $g=1$. This is both because the expressions can then be expressed in a more compact form, and also 
because it allows us to highlight a novel feature for the $g=1$ case.

\subsubsection*{Genus $g\neq 1$}

In this case it is convenient to rescale the fluxes $M_a$ as well 
as the flavour flux parameter $n_2$ as follows:
\begin{align}\label{Mandn}
M_a \ \equiv \ m_a (g-1)N~,\qquad n_2\ =\ n_3\ \equiv \ \pt (g-1)\,,\qquad \text{for $g\ne 1 $}\,.
\end{align}
Using \eqref{Ypqv} we can immediately obtain an explicit formula for
master volume function
$\mathcal{V}(\vec{b};\{\lambda_a\})$ given in \eqref{mvolfn} and hence the various derivatives
appearing in the expressions \eqref{constraintformula}--\eqref{Ssusyformula}, taking care to set $b_1=2$ after taking the derivatives. We can then solve the equations as follows.
Start with the constraint condition \eqref{constraintformula}, the expressions for the flux $N$ given
in \eqref{Nformula} and one of the fluxes $M_a$ given in \eqref{Maformula}, which we take to be $M_1$ for definiteness. We can then use these to solve for $A$ and two of the four $\lambda_a$, 
say $\lambda_3, \lambda_4$, in terms of $N,m_1$, $\pt$, $b_2$ , $b_3$
as well as $\lambda_1, \lambda_2$. Since there are only two independent K\"ahler class
parameters, it must be the case, and indeed it is, that 
$\lambda_1$ and $\lambda_2$ will drop out of any final formula.
The remaining three fluxes $m_a$ (giving the $M_a$) for $a=2,3,4$ are then given by
\begin{align}\label{Mandn2}
m_2 \ =\ m_4\ = \ \frac{-{m} p+2 p+{\pt}}{p+q}\,,\qquad
m_3\ =\ \frac{({m}-2) (p-q)-2 {\pt}}{p+q}\,,
\end{align}
where we have defined $m\equiv m_1$. The equality of $m_2$ and $m_4$ arises because
our twisting preserves $SU(2)$ symmetry.
One can immediately check that these satisfy the 
relation \eqref{Marel} arising from the homology relation \eqref{homrel}, as expected.

The action, $\Ssusy$, given in \eqref{Ssusyformula} and hence the trial central charge, $\cZ$, given by
\eqref{cS24}, can now easily be calculated. As the form is rather long, it is more conveniently
written after performing the following linear change of variables
\bea\label{newvars}
b_2 &=& (p-q-2)\epsilon_1 -p\epsilon_2 + p-q~, \qquad b_3 \ = \ (p-q)\epsilon_1  -p\epsilon_2 + p-q~.
\eea
In fact we will see later in this subsection that this change of variables can actually be {\it derived} from the field theory analysis, where $\epsilon_1$ and $\epsilon_2$ are field theory variables. 
In the 
new $(\epsilon_1,\epsilon_2)$ variables the trial central charge (after setting $b_1=2$) is given by
\bea\label{cZYpq}
&&\cZ(\epsilon_1,\epsilon_2)\ =\  6(g-1)N^2\Big\{\left[ \pt+(p-q)(1-m)\right]\epsilon_1^2\nn\\
&& \qquad \qquad \quad + p\tfrac{\pt^2 \left(p^2+p q+q^2\right)+\pt p \left[p^2+m p q-(m-3) q^2\right]+p^2 \left[(m-1) (p^2+2pq)+(3+(m-3) m) q^2\right]}{(p+q)^2 \left(p^2-(\pt+p) q\right)} \epsilon_2^2\nn\\
&& \qquad \qquad \quad +2\tfrac{ p^2 (\pt+p) (\pt+p-m p)-p^2 \left[\pt (3-2 m)+(m-2)^2 p\right] q+\left[\pt^2-\pt (m-3) p+(3+(m-3) m) p^2\right] q^2}{(p+q) \left(p^2-(\pt+p) q\right)}\epsilon_2\nn\\
&&\qquad \qquad \quad + \tfrac{[3+(m-3) m] p^2 (p-q)^2-\pt p (p-q) \left[(2 m-3) p-(m-3) q\right]+\pt^2 \left(p^2-p q+q^2\right)}{p^3-p (\pt+p) q}\Big\}~.
\eea
In particular, notice there is no $\epsilon_1$ term in this quadratic, which immediately sets $\epsilon_1=0$ for the critical point. From (\ref{newvars}) this then implies $b_2=b_3$ at the critical point, which was expected due to the fact the twist preserves $SU(2)$ symmetry.
The critical $R$-symmetry vector has $\vec{b}=(2,b_2,b_2)$ with
\begin{align}
\textstyle{
b_2 \ =\ p\frac{- p^3 \left[({m}-3) {\pt}+({m}-2)^2 q\right]+p^2 \left[2 ({m}-1) {\pt} q+({m}-2)^2 q^2+2 {\pt}^2\right]+p {\pt} q ({\pt}-({m}-3) q)+{\pt}^2 q^2}{({m}-1) p^4+p^3 \left[2 ({m}-1) q+{\pt}\right]+p^2 \left[{m} {\pt} q+(({m}-3) {m}+3) q^2+{\pt}^2\right]+p {\pt} q ({\pt}-({m}-3) q)+{\pt}^2 q^2}\,.
}
\end{align}
Furthermore, the on-shell central charge, $\csugra=\cZ  |_\mathrm{on-shell}$, is given by
\begin{equation}
\label{cchgegenypq}
{\textstyle
\csugra \ = \ 
\frac{6 (g-1) {N}^2 p (({m}-2) p-{\pt}) \left[({m}-2) ({m}-1) p^3+p^2 (-2 {m} {\pt}-({m}-2) ({m}-1) q+{\pt})+({m}-3) p {\pt} q-{\pt}^2 q\right]}{({m}-1) p^4+p^3 (2 \left[({m}-1) q+{\pt}\right]+p^2 \left[{m} {\pt} q+(({m}-3) {m}+3) q^2+{\pt}^2\right]+p {\pt} q ({\pt}-({m}-3) q)+{\pt}^2 q^2}\,,
}
\end{equation}
and the on-shell $R$-charges are 
\bea\label{rchgegenypq}
R_1&=&
{\textstyle
-\frac{{N} (p+q) \left[({m}-4) ({m}-1) p^3-p^2 ((3 {m}-4) {\pt}+(({m}-3) {m}+4) q)+p {\pt} (({m}-4) q+{\pt})-{\pt}^2 q\right]}{({m}-1) p^4+p^3 (2 ({m}-1) q+{\pt})+p^2 \left({m} {\pt} q+(({m}-3) {m}+3) q^2+{\pt}^2\right)+p {\pt} q ({\pt}-({m}-3) q)+{\pt}^2 q^2}\,,
}
\nonumber\\[1.5mm]
R_2\ =\ R_4&=&
{\scriptstyle
\frac{{N} p^2 (({m}-2) p-{\pt}) (({m}-1) p-{\pt}+q)}{({m}-1) p^4+p^3 (2 ({m}-1) q+{\pt})+p^2 \left({m} {\pt} q+(({m}-3) {m}+3) q^2+{\pt}^2\right)+p {\pt} q ({\pt}-({m}-3) q)+{\pt}^2 q^2}\,,
}
\nonumber\\[1.5mm]
R_3&=&
{\scriptstyle
-\frac{{N} (p+q) \left[({m}-2) ({m}-1) p^3-p^2 ({m} {\pt}+({m}-2) ({m}-1) q)-p {\pt} ({\pt}-({m}-2) q)-{\pt}^2 q\right]}{({m}-1) p^4+p^3 (2 ({m}-1) q+{\pt})+p^2 \left({m} {\pt} q+(({m}-3) {m}+3) q^2+{\pt}^2\right)+p {\pt} q ({\pt}-({m}-3) q)+{\pt}^2 q^2}~.
}\eea

\subsubsection*{Genus $g=1$, $n_2\neq 0$}

With the twisting still taken to preserve $SU(2)$, so that  $n_2=n_3$, we begin by rescaling the fluxes
via 
\begin{align}
M_a\ =\  m_aN\,,\qquad \text{for $g=1$}\,.
\end{align}
When $n_2\ne 0$ the procedure is essentially identical to the $g\neq 1$ case above, and we again just record the final results.
The fluxes $m_a$ for $a=2,3,4$ are given by
\begin{align}
m_2\ =\ m_4\ =\ \frac{n_2-m p}{p+q}
\,,\qquad
m_3\ =\ \frac{m (p-q)-2 n_2}{p+q}\,,
\end{align}
with $m\equiv m_1$, which clearly satisfy  \eqref{Marel}.
Using the new variables given in \eqref{newvars}
the trial central charge (after setting $b_1=2$) is given by
\begin{align}\label{cZYpqg1}
&\cZ(\epsilon_1,\epsilon_2)\ =\  6N^2\Big\{\left[ m (q-p)+n_2\right]\epsilon_1^2\nn\\
& \qquad \qquad 
-\frac{ p \left(m^2 p^2 q^2+m n_2 p q (p-q)+n_2^2 \left(p^2+p q+q^2\right)\right)}{n_2 q (p+q)^2}
\epsilon_2^2\nn\\
& \qquad \qquad +2
\frac{ \left(m^2 p^2 q (p-q)+m n_2 p (p-q)^2-n_2^2 \left(p^2+q^2\right)\right)}{n_2 q (p+q)}
\epsilon_2\nn\\
&\qquad \qquad
-\frac{ \left(m^2 p^2 (p-q)^2-m n_2 p \left(2 p^2-3 p q+q^2\right)+n_2^2 \left(p^2-p q+q^2\right)\right)}{n_2 p q}
\Big\}~.
\end{align}
The critical $R$-symmetry vector has $\vec{b}=(2,b_2,b_2)$ with
\begin{align}
b_2\ =\ \frac{p \left[m^2 p^2 q (q-p)-m n_2 p (p-q)^2+n_2^2 \left(2 p^2+p q+q^2\right)\right]}{m^2 p^2 q^2+m n_2 p q (p-q)+n_2^2 \left(p^2+p q+q^2\right)}\,,
\end{align}
and the on-shell central charge, $\csugra=\cZ  |_\mathrm{on-shell}$, is given by
\begin{align}
\csugra\ =\ \frac{6 N^2 p (m p-n_2) \left[m^2 p^2 (p-q)+m n_2 p (q-2 p)-n_2^2 q\right]}{m^2 p^2 q^2+m n_2 p q (p-q)+n_2^2 \left(p^2+p q+q^2\right)}\,.
\end{align}
The on-shell $R$-charges are 
\begin{align}\label{Rn2}
R_1&\ =\ -\frac{N (p+q) \left[(m^2 p^2 +n_2^2)(p-q)+m n_2 p (q-3 p)\right]}{m^2 p^2 q^2+m n_2 p q (p-q)+n_2^2 \left(p^2+p q+q^2\right)}\,,\nonumber\\[1.5mm]
R_2\ =\ R_4&\ =\ \frac{N p^2 (n_2-m p)^2}{m^2 p^2 q^2+m n_2 p q (p-q)+n_2^2 \left(p^2+p q+q^2\right)}\,,\nonumber\\[2.2mm]
R_3&\ =\ -\frac{N (p+q) \left[m p(p-q)(mp-n_2)-n_2^2 (p+q)\right]}{m^2 p^2 q^2+m n_2 p q (p-q)+n_2^2 \left(p^2+p q+q^2\right)}\,.
\end{align}

\subsubsection*{Genus $g=1$, $n_2=0$}

 Interestingly, for this particular case we now need to proceed slightly
differently.\footnote{We note that for  $p>q>0$
this is also an example of an obstructed AdS$_3\times Y_7$ geometry \cite{Couzens:2018wnk}, of the type discussed
in section \ref{sec:exist}.  Specifically, the critical $R$-symmetry vector, given in \eqref{critrgz}, lies outside the Reeb cone. We
discuss this further at the beginning of section \ref{secexplexamples}.} In this case the constraint equation \eqref{constraintformula} is independent of the $\lambda_a$ and so it must
be solved for one of the components $b_2,b_3$, and we choose to solve it for $b_3$. 
The expressions for the fluxes $M_a$ are also independent of the $\lambda_a$. Using the expression
for $M_1$ we can solve for $A$. Finally, we can use the expression for $N$ to solve for one of the
$\lambda_a$, which we choose to be
$\lambda_4$.
The fluxes $M_a$ for $a=2,3,4$ are then given by
\begin{align}
M_2\ =\ M_4\ =\ -\frac{M_1 p}{p+q}
\,,\qquad
M_3\ =\ \frac{M_1( p- q)}{p+q}\,.
\end{align}
Clearly the condition \eqref{Marel} is satisfied.
The off-shell central charge, with $b_1=2$, can be written as
\begin{align}\label{conlyb2}
\cZ\ =\ \frac{6 M_1 N (q-p) (b_2 q+2 p) [b_2 q+2 p (p-q-1)]}{q^2 (-p+q+2)^2}\,,
\end{align}
which is a quadratic in $b_2$ only (as we have already solved for $b_3$).
Extremizing this with respect to $b_2$, and combining with the previously obtained expression
for $b_3$, we find that the critical $R$-symmetry vector has
$\vec{b}=(2,b_2,b_2)$ with
\begin{align}\label{critrgz}
b_2\ =\ \frac{p(q-p)}{q}\,,
\end{align}
and the on-shell central charge, $\csugra = \cZ  |_\mathrm{on-shell}$, is given by
\begin{align}
\csugra\  =\ \frac{6 M_1 N p^2 (p-q)}{q^2}\,.
\end{align}
Furthermore, and interestingly, the on-shell $R$-charges can be written in the form
\begin{align}\label{Rn2zero}
R_1\ &=\ 
\frac{N (q^2-p^2)}{q^2}+(p+q)N\gamma~,
\nonumber\\
R_2\ =\ R_4\ &=\ \frac{N p^2}{q^2}-pN\gamma~,
\nonumber\\
R_3&\ =\ \frac{N (q^2-p^2)}{q^2}+(p-q)N\gamma~,
\end{align}
where $\gamma$ is an undetermined parameter given by
\begin{align}
\gamma\ =\ 
\frac{p}{q^2}-\frac{L^4 q [\lambda_1 (p+q)+\lambda_3 (q-p)]}{2 \pi  N g_s \ell_s^4 p^2 (p^2-q^2)}\,.
\end{align}
Since $\lambda_4$ was already fixed, we can view $\gamma$ as parametrizing an undetermined
transverse K\"ahler class. In other words, for this particular case, with
$g=1$ and $n_2=0$, the $c$-extremization procedure that we are implementing does not fix one
of the two independent transverse K\"ahler classes. 
This was the novel feature concerning this case that we wanted to highlight.

To clarify this feature further, we first note that we can recover the results for the central
charge and fluxes given in \cite{Couzens:2018wnk}. These quantities were calculated
in \cite{Couzens:2018wnk} both using an explicit 
supergravity solution, confirming the results of \cite{Donos:2008ug},
as well as using a localization formula. We also recall that the supergravity solutions
only exist for $q>p>0$, for which the polyhedral cone $\mathcal{C}$ associated with the $\v_a$ in \eqref{Ypqv} 
is \emph{not convex} (see the discussion at the beginning of section 7).
Specifically, if we set
$M_1=-(p+q)M$ in the above expressions we get
\begin{align}\label{emmsspc}
M_2\ =\ M_4\ =\ pM,\qquad
M_3\ =\ (q-p)M,
\end{align}
and
\begin{align}\label{csugran2zg1}
\csugra\ =\ \frac{6 M N p^2 (q^2-p^2) }{q^2}~,
\end{align}
in agreement with \cite{Couzens:2018wnk}. The $R$-charges that were given in \cite{Couzens:2018wnk} were obtained from
the explicit supergravity solution. In order to recover the expressions in \cite{Couzens:2018wnk}
we need to impose the extra condition 
that $\gamma=0$, or equivalently
\begin{align}
\lambda_3\ =\  (p+q) \left(\frac{\lambda_1}{p-q}-\frac{2 \pi  g_s \ell_s^4 N p^3}{L^4 q^3}\right)\,.
\end{align}
At this point one might conjecture that supergravity solutions generalising
the one discussed in \cite{Couzens:2018wnk}, with an extra parameter
associated with the free transverse K\"ahler class, might exist. 
However, we do not think this is the case. We first point out that
$\gamma=0$ arises \emph{naturally} in (\ref{Rn2zero}) as the $n_2\rightarrow 0$ of 
the $R$-charges (\ref{Rn2}) with general $n_2$.
Secondly, as we shall discuss further at the very end of the subsection, the associated field theory
analysis clearly shows that $\gamma=0$.

\subsubsection*{Dual field theory and $c$-extremization}

We now turn attention to the $c$-extremization procedure in field theory, starting with
the quiver gauge theory dual to the $Y^{p,q}$  Sasaki-Einstein spaces \cite{Benvenuti:2004dy}, which have $p>q>0$. 
The field
content of these theories is presented in Table \ref{table1}. The gauge group is $SU(N)^{2p}$, 
the $\lambda$ are the gauginos, and the remaining fields are bifundamental matter fields. 
The $U(1)_{B}$ corresponds to the baryonic symmetry associated to the single non-trivial three-cycle of $Y^{p,q}$, while $U(1)_{F_i}$, $i=1,2$, are flavour symmetries corresponding to $U(1)$ isometries 
under which the holomorphic volume form $\Omega_{(3,0)}$ is uncharged. In particular $U(1)_1\subset SU(2)$ is the Cartan of the $SU(2)$ isometry that acts on the round $S^2$ in the metric. We emphasize that
$R_0$ is \emph{not} the $R$-charge of the dual 
SCFT in $d=4$ (which can be found in \cite{Benvenuti:2004dy}). Instead, as in \cite{Couzens:2018wnk}, $R_0$ is a simple fiducial $R$-charge that can be used in the $c$-extremization procedure for the putative $d=2$ SCFT. Geometrically, $R_0$ corresponds to the Killing 
vector $\partial_\psi$ in the $Y^{p,q}$ metric, in the original coordinates used in \cite{Gauntlett:2004yd}.
\begin{table}[h!]
\small{
\begin{center}
\begin{tabular}{|c|c|c|c|c|c|}
\hline
Field&Multiplicity&$R_0$-charge&$U(1)_{B}$& $U(1)_{F_{1}}$&$U(1)_{F_{2}}$\\
\hline\hline
$Y$&$(p+q) N^2$&$0$&$p-q$&$0$&$-1$\\
$Z$&$(p-q) N^2$&$0$&$p+q$&$0$&$1$\\
$U_1$&$p N^2$&$1$&$-p$&$1$&$0$\\
$U_2$&$p N^2$&$1$&$-p$&$-1$&$0$\\
$V_1$&$q N^2$&$1$&$q$&$1$&$1$\\
$V_2$&$q N^2$&$1$&$q$&$-1$&$1$\\
$\lambda$&$2p(N^2-1)$&1&0&0&0\\
\hline
\end{tabular}
\caption{The field content of the $Y^{p,q}$ quiver theories.}
\label{table1}\end{center}
}
\end{table}

We consider these $d=4$ SCFTs theories wrapped on $\Sigma_g$, with a partial topological twist given by a background gauge field switched on along the generator
\bea\label{Ypqtop}
T_{\mathrm{top}} & =&  f_2T_{F_2}+\betabeta T_{B}+\frac{\kappa}{2}R_{R_0}~.
\eea
Here $T_{F_2}$, $T_B$ and $T_{R_0}$ are the generators of $U(1)_2$, $U(1)_B$ and the fiducial $R$-symmetry, respectively, and
\bea
\kappa &= & \begin{cases}\ 1 & \quad g \, = \, 0~, \\ \ 0 & \quad g \, = \, 1~, \\ \ -1 & \quad g\, >\, 1~.\end{cases}
\eea
Notice that generically the four-dimensional superconformal $R$-symmetry is a linear combination of $R_0$ \emph{and} 
$U(1)_B$, $U(1)_{F_2}$. Our basis is hence different to that used in \cite{Benini:2015bwz}, implying that
the parameters $f_2$, $B$ in (\ref{Ypqtop}) are generically different to the analogous parameters appearing in \cite{Benini:2015bwz}.
The fact that $T_{F_1}$ does not appear in \eqref{Ypqtop} is precisely the condition that we preserve the
$SU(2)$ flavour symmetry.
The trial $R$-charge is a linear combination
\bea\label{YpqtrialR}
T_{\text{trial}} &=&  T_{R_0}+\epsilon_BT_{B}+ \epsilon_{1} T_{F_{1}}+\epsilon_{2} T_{F_{2}}~,
\eea
where $\epsilon_B$, $\epsilon_i$ are parameters. The trial $c$-function is 
given by \cite{Benini:2012cz}
\begin{align}\label{ceetrialgenexp}
c(\epsilon_B,\epsilon_1,\epsilon_2)\ =\ -3\bar\eta \sum_\sigma m_\sigma t_\sigma (q_{R}^{(\sigma)})^2\,,
\end{align}
where the sum is over all fermion fields, labelled by $\sigma$, $m_\sigma$ is their multiplicity, $t_\sigma$ is the charge
under the background gauge field \eqref{Ypqtop} and $q_{R}^{(\sigma)}$ is the charge with respect to the trial R-symmetry
\eqref{YpqtrialR}. In addition
\bea\label{etadef2}
\bar\eta &\equiv & \begin{cases}\ 2|g-1| & \quad g\, \neq \, 1~,\\ \  2 & \quad g\, = \, 1~,\end{cases}
\eea
and we note that $\bar\eta\kappa=2(1-g)$ for all $g$.
Using Table \ref{table1} we find
\bea\label{cYpqfieldtheory}
c(\epsilon_B,\epsilon_1,\epsilon_2) & =& -3 \bar\eta  N^2 \Big\{2 B \left[ p^2 \left(2 q \epsilon_2 \epsilon_B-\epsilon_1^2+\epsilon_2^2+1\right)+2 p (q^2-p^2) \epsilon_B+q^2 \left(\epsilon_1^2-1\right)\right]\nn\\
&& +2 f_2 q \left(p^2 \epsilon_B^2+\epsilon_1^2-1\right)+4 f_2 p \epsilon_2 (p \epsilon_B-1)+\kappa  q^2 \epsilon_B (p \epsilon_B-2)\nn\\
&& -\kappa  p \left(p \epsilon_B (p \epsilon_B-2)+\epsilon_2^2\right)-2 \kappa  q \epsilon_2\Big\}~.
\eea

The $c$-extremization procedure requires us to find the critical point of the quadratic (\ref{cYpqfieldtheory}) in $(\epsilon_B,\epsilon_1,\epsilon_2)$. 
However, we shall do this in two stages. We first extremize  (\ref{cYpqfieldtheory})
 over the baryonic mixing parameter $\epsilon_B$, which gives
\bea\label{extremizeeB}
\frac{\partial c}{\partial \epsilon_B} \ =\  0 \quad \Longrightarrow \quad \epsilon_B \ = \ \frac{1}{p}-\frac{2 \left[B \left(p q \epsilon_2+q^2-p^2\right)+f_2 (p \epsilon_2+q)\right]}{2 f_2 p q+\kappa  \left(q^2-p^2\right)}~.
\eea
Substituting this result back into (\ref{cYpqfieldtheory}) then gives
\begin{align}
\label{ctwoYpq}
 c(\epsilon_1,\epsilon_2) \ = \  -3\bar\eta N^2 \Big\{&2 \left[B \left(q^2-p^2\right)+f_2 q\right]\epsilon_1^2 + p \Big[ \frac{4 p^2 (B q+f_2)^2}{\kappa  (p^2-q^2) -2 f_2 p q}+ 2 B p-\kappa\Big]\epsilon_2^2 \nn\\
& +\Big[4 B p q-2 \kappa  q -\frac{8 p^2 (B q+f_2) (B (p^2-q^2) -f_2 q)}{\kappa  (p^2-q^2) -2 f_2 p q}\Big]\epsilon_2\nn\\
&+\frac{4 p \left(B \left(q^2-p^2\right)+f_2 q\right)^2}{\kappa  (p^2-q^2) -2 f_2 p q}-2 B (p^2-q^2) +\frac{\kappa  (p^2-q^2)}{p}\Big\}~.
\end{align}

\subsubsection*{Genus $g\neq 1$}

For $|\kappa|=1$, let us compare (\ref{ctwoYpq}) to the corresponding $g\neq 1$ supergravity trial central charge function 
 (\ref{cZYpq}). 
Remarkably, after making the linear change of variable (which we come back to below)
\bea\label{Bf2}
B&=& \frac{p(m-1)+q-\pt }{2 p (p+q)}\kappa~, \qquad f_2 \ =\  \frac{\pt +p-q}{2p}\kappa~,
\eea
the field theory trial $c$-function (\ref{ctwoYpq}) precisely matches the supergravity trial central charge function 
(\ref{cZYpq}). Thus after extremizing over the baryon mixing parameter $\epsilon_B$, the field theory and gravity 
$c$-functions agree \emph{off-shell}! Of course, it is then immediate that the central charges 
computed on both sides will agree. Note this is true for an arbitrary $Y^{p,q}$ fibred 
over an arbitrary genus $g\neq 1$ Riemann surface, with arbitrary flavour twisting parameter $f_2$ 
and baryon flux $B$.
The change of variables in  (\ref{newvars}) and the second equation in (\ref{Bf2}) is simply because the basis for the $U(1)^2$ (non $R$-symmetry) flavour symmetries we used in the geometry computation is 
different to that in Table \ref{table1}. Similarly, the parameter $m$ in the geometry computation 
was defined as $m=M_1/(N(g-1))$, where $M_1$ is the quantized five-form flux through the first toric divisor. 
This is then necessarily linearly related to the parameter $B$ in field theory, which is instead associated to the 
flux through the generating three-cycle of $Y^{p,q}$. 

Even more remarkably, the \emph{off-shell} $R$-charges also agree. Before extremizing over $\epsilon_B$, the trial $R$-charges 
of the fields $(X_1,X_2,X_3,X_4)\equiv (Z,U_2,Y,U_1)$ are respectively (see Table \ref{table1})
\begin{align}
R[X_1] \ & = \ \epsilon_2 +(p+q)\epsilon_B~, \qquad \quad R[X_2]\ = \ 1 - \epsilon_1 - p\epsilon_B~, \nn\\ \quad R[X_3] \ & =  \ -\epsilon_2 + (p-q)\epsilon_B~, \qquad 
 R[X_4]\ = \ 1 + \epsilon_1 - p\epsilon_B~.
\end{align}
Extremizing the trial $c$-function over $\epsilon_B$ gives (\ref{extremizeeB}). Substituting in for the critical value of 
$\epsilon_B$ then gives the $R$-charges
\bea\label{RYpqoff}
R[X_1] &=& \epsilon_2+\frac{(p+q) \left[2 p \left(f_2  p \epsilon_2+B \left(-p^2+q^2+p q \epsilon_2\right)\right)+(p^2-q^2) \kappa \right]}{p [-2 f_2 p q+(p^2-q^2) \kappa ]}~,\nn\\
R[X_2] &=&-\epsilon_1 +\frac{2 p \left[f_2 (q+p \epsilon_2)+B \left(-p^2+q^2+p q \epsilon_2\right)\right]}{2 f_2 p q+\left(-p^2+q^2\right) \kappa }~,\nn\\
R[X_3]&=& -\epsilon_2+\frac{(p-q) \left[2 p \left(f_2  p \epsilon_2+B \left(-p^2+q^2+p q \epsilon_2\right)\right)+(p^2-q^2) \kappa \right]}{p [-2 f_2 p q+(p^2-q^2)  \kappa ]}~,\nn\\
R[X_4] &=&\epsilon_1 +\frac{2 p \left[f_2 (q+p \epsilon_2)+B \left(-p^2+q^2+p q \epsilon_2\right)\right]}{2 f_2 p q+\left(-p^2+q^2\right) \kappa }~,
\eea
as functions of the remaining trial $R$-charge parameters $(\epsilon_1,\epsilon_2)$.
After the change of variable  (\ref{newvars}), (\ref{Bf2}), remarkably these functions agree with the geometric $R$-charges, namely 
\bea\label{compareRs}
R_a \ = \ R[X_a]N~, \qquad a=1,\ldots,4~,
\eea
where $R_a$ are defined in (\ref{rchgegenexpgeneral}). Here, as for the trial central charge, we have 
imposed the constraint and flux quantization conditions in the geometry computation, so that the resulting trial $R$-charges (\ref{rchgegenexpgeneral})
are functions of the $R$-symmetry vector $\vec{b}=(2,b_2,b_3)$, or equivalently functions of the parameters $(\epsilon_1,\epsilon_2)$ introduced in 
(\ref{newvars}). Physically, the gauge-invariant baryonic operator $\det X_a$ constructed from each of the fields $X_a$ 
has $R$-charge $R[X_a]N$, and is dual to a D3-brane wrapped on the corresponding toric 
three-submanifold $S_a$, as in 
(\ref{compareRs}). Since the $R$-charges match off-shell, they of course also match on-shell.

We now return to the change of the geometric variables $(b_2,b_3)$ to
the field theory variables $(\epsilon_1,\epsilon_2)$ that we introduced in \eqref{newvars}. This may be derived from 
field theory, as follows. 
Consider the expressions for the $R$-charges \eqref{RYpqoff},
obtained in field theory after substitution of the critical value of $\epsilon_B$. Using 
 the toric data $v_a^i$ on the geometry side we can calculate
(see \eqref{summarsv})
\bea\label{togetbeprel}
\sum_{a=1}^d v_a^i R[X_a] &=& \ b_i ~.
\eea
On the left hand side the $R[X_a]$ are given in (\ref{RYpqoff}), which are functions of $\epsilon_1,\epsilon_2$, and (\ref{togetbeprel})
precisely implements the change of variable \eqref{newvars}.

We also note that the linear change of variables given in \eqref{Bf2} satisfies the following simple relation
\begin{align}\label{simprelMtee}
M_a\ =\ -\bar\eta\, t[X_a] N\,.
\end{align}
Here the $M_a$ are the fluxes on the gravity side, given for the $Y^{p,q}$ case in 
\eqref{Mandn}, \eqref{Mandn2}, while the $t[X_a]$ are the charges\footnote{Note that in \eqref{ceetrialgenexp} the charges,
$t_\sigma$, of the fermion fields with respect to \eqref{Ypqtop} appeared.} of the bosonic fields
$(X_1,X_2,X_3,X_4)\equiv (Z,U_2,Y,U_1)$ of the quiver gauge theory with respect to the background gauge field given
in \eqref{Ypqtop}. We conjecture that this is a general result, and we will see that it is satisfied
for all of the examples in this paper. For example, for the universal twist, considered in the last section, the relation
\eqref{MaRa} can immediately be written in the form \eqref{simprelMtee}.

\subsubsection*{Genus $g=1$}

Finally, it is straightforward to derive similar results in the genus $g=1$ case, where equivalently $\kappa=0$. In the case that $f_2\ne 0$,
everything works in the same way, with the change of variable (\ref{Bf2}) replaced by 
\bea
B&=& \frac{n_2-mp}{2p(p+q)}~, \qquad f_2 \ =\  -\frac{n_2}{2p}~,
\eea
which again satisfies \eqref{simprelMtee}.
The off-shell trial $c$-function in field theory (\ref{ctwoYpq}), with $\kappa=0$, then matches the gravity trial $c$-function (\ref{cZYpqg1}), 
with the $\kappa=0$ $R$-charges (\ref{RYpqoff}) similarly matching the geometric $R$-charges  (\ref{rchgegenexpgeneral}), as in (\ref{compareRs}).

For the $g=1$ case with $f_2= 0$ we need to treat the field theory calculation slightly differently,
mirroring to some extent what we saw in the geometric calculation (just below equation \eqref{Rn2}). We can no longer
solve for $\epsilon_B$ as we did in \eqref{extremizeeB}. Instead, if we extremize 
the central charge \eqref{cYpqfieldtheory} with respect to $\epsilon_i$ and $\epsilon_B$ we
find 
\begin{align}
\epsilon_1\ =\ 0~,\qquad
\epsilon_2\ =\ \frac{p^2-q^2}{pq}~,\qquad
\epsilon_B\ =\ \frac{q^2-p^2}{pq^2}~.
\end{align}
If we now set $B= -\frac{M_1}{2(p+q)N}$ and $f_2=0$, consistent with \eqref{simprelMtee},
then we find that the on-shell central charge agrees 
with the on-shell geometric result given in \eqref{csugran2zg1} (after setting $M_1=-(p+q)M$). Furthermore, for the $R$-charges using the identification
as in \eqref{compareRs} we find agreement with \eqref{Rn2zero}, provided that we set $\gamma=0$.

On the other hand, we can perform the $c$-extremization slightly differently in this case, as follows. Since
\bea
\frac{\partial c}{\partial \epsilon_B} &=& 24B N^2 p\left(p^2-q^2-pq\epsilon_2\right)~,
\eea
extremizing over the baryon mixing necessarily sets
\bea\label{e2sol}
\epsilon_2 &=& \frac{p^2-q^2}{pq}~.
\eea
Substituting this back into the trial central charge then gives
\bea\label{conly1}
c(\epsilon_B,\epsilon_1) &=& \frac{12 BN^2(q^2-p^2)(p^2-q^2\epsilon_1^2)}{q^2}~.
\eea
Notice this is  independent of $\epsilon_B$, so we have ``lost'' the extremal equation for this baryon mixing parameter! 
This is simply a consequence of the fact that in this case $\epsilon_B$ appears linearly in $c(\epsilon_B,\epsilon_1,\epsilon_2)$,
and thus extremizing over it and substituting back into $c$ sets its coefficient to zero.
One easily verifies (\ref{conly1}) agrees with the off-shell gravity result (\ref{conlyb2}), where as above we set 
$B= -\frac{M_1}{2(p+q)}$, and change variable from $\epsilon_1$ to $b_2$ using \eqref{e2sol} and (\ref{newvars}). 
Extremizing (\ref{conly1}) sets $\epsilon_1=0$, leaving $\epsilon_B$ free. The $R$-charges of the fields are then
\bea\label{Rspecial}
R[Z] &=& \frac{q^2-p^2}{q^2} + (p+q)\gamma~,\nn\\
R[U_1]\ =\ R[U_2] &=& \frac{p^2}{q^2}-p\gamma~,\nn\\
R[Y] &=& \frac{q^2-p^2}{q^2} + (p-q)\gamma~,
\eea
where we have substituted
\bea
\epsilon_B &=& \frac{q^2-p^2}{pq^2}+\gamma~.
\eea
The $R$-charges (\ref{Rspecial}) agree with (\ref{Rn2zero}). 

The above discussion makes it clear, in the context of the field theory analysis, that the apparently unconstrained parameter $\gamma$ in (\ref{Rspecial}), that we saw in the gravitational calculation, is an artefact 
of only partially performing the extremization: extremizing over a parameter that appears linearly, and then substituting back into 
$c$, will miss the equation of motion for that parameter, leaving it unconstrained. In
field theory we have additional baryon mixing parameters, $\epsilon_B$, which must be extremized over before matching to the off-shell gravity central charge function, and this sets $\epsilon_B\ =\ \frac{q^2-p^2}{pq^2}$ and hence $\gamma=0$.
This suggests there might be a more general way to go off-shell in gravity, at least in the special cases where the equation of motion $\gamma=0$ 
arises directly from this modified extremal problem, rather than as a limit of the general $n_2$ equations discussed earlier in the section.

\subsection{$X^{p,q}$}\label{sec:dp2example}

Having illustrated the general procedure in detail for the $Y^{p,q}$ spaces, it is now straightforward to implement our extremal 
problem in gravity for any choice of toric Calabi-Yau 3-fold singularity. After specifying the toric data, given by the inward pointing normal vectors 
$\v_a$, the process is then entirely algorithmic. In this section we briefly outline the steps and key formulae 
for the case when $Y_5=X^{p,q}$. We will present most of the formulae for general $p>q>0$, but 
then specialise to the case of $X^{2,1}$ for some of the more unwieldy expressions. Here $X^{2,1}$ is 
the $SE_5$ manifold associated to the 
canonical complex cone over the second del Pezzo surface, $dP_2$. Note that 
 while there are also explicit supergravity solutions for some of the $Y^{p,q}$ cases, in the $X^{p,q}$ case the 
equation of motion (\ref{boxReqn}) is genuinely a PDE in 2 variables, with no reason to expect it to separate variables 
into decoupled ODEs, and thus it seems very unlikely explicit supergravity solutions can be constructed in this 
case. Our gravitational $c$-extremization is hence the only method available to compute the central charge and $R$-charges 
on the gravity side, assuming the solution exists. 

The toric Calabi-Yau cones for the $X^{p,q}$ spaces have
 inward pointing normal vectors
\bea\label{Xpqv}
\v_1 &=& (1,1,0)~, \quad \v_2 \ = \ (1,2,0)~, \quad \v_3 \ = \ (1,1,p)~, \N\\
 \v_4 & = & (1,0,p-q+1)~, \quad \v_5 \ = \ (1,0,p-q)~.
\eea
The flavour twisting parameters are $\vec{n}=(2-2g,n_2,n_3)$, where we keep $n_2,n_3\in\Z$ general. 
Again, for illustrative purposes we only present formulae for the generic $g\neq 1$ case. 
As in the previous section we rescale the 
 fluxes $M_a$, $a=1,\ldots,5$, and 
the flavour flux parameters $n_2, n_3$ as follows:
\begin{align}\label{mandnxpq}
M_a \ \equiv \ m_a  (g-1)N,\qquad n_i\ \equiv \ \ \pt_i (g-1)\,,\qquad i\, =\, 2,3\,.
\end{align}
After computing the master volume function \eqref{mvolfn}, as before we then solve the 
 constraint condition \eqref{constraintformula}, the expression for the flux $N$ given
in \eqref{Nformula}, and two of the fluxes $M_a$ given in \eqref{Maformula}, which we take to be $M_1, M_2$ for definiteness. 
We can then use these to solve for $A$ and three of the five $\lambda_a$, in terms of $N,m_1,m_2,\pt_2,\pt_3,b_2,b_3$. 
The remaining two K\"ahler class parameters then drop out of the remaining formulae, as they must. 
The remaining three fluxes $m_a$ (giving the $M_a$) for $a=3,4,5$ are then given by
\bea\label{mandnxpq2}
m_3 &=& -m_1 - 2 m_2 - \pt_2~, \quad 
m_4\ =\  p (m_1+m_2-2)+q (m_2+\pt_2+2)-\pt_3~,
\nn\\  
m_5 & = & p (-m_1-m_2+2)-q (m_2+\pt_2+2)+m_2+\pt_2+\pt_3+2\,.
\eea
One can immediately check that these satisfy the 
relation \eqref{Marel}, as expected.

It is again convenient to make a linear change of variables
\bea\label{beetoepdp2}
b_2 &=& 2+\epsilon_1~,\qquad b_3 \ = \  2p+\epsilon_2~,
\eea
which, as in the last subsection, will be derived later using the field theory analysis. 
The expression for the trial central charge function in gravity for general $p$ and $q$ is a little too long 
to present here, so instead we present the result for $X^{2,1}$:
\bea\label{cdP2grav}
&& \cZ(\epsilon_1,\epsilon_2)\ =\  6(g-1)N^2\Big\{\tfrac{-16 (m_1+m_2-2)^2+8 s_2 s_3+2 [-2 (m_1+m_2-4)+s_2] s_3^2+3 s_3^3}{2 (8 s_2+s_3 (8+s_3))}\epsilon_1^2\nn\\
&& + \tfrac{2 \left[m_1^2 (s_2-2)+2 m_1 s_2 (2+m_2+s_2)+s_2 \left(2 s_2+(m_2+s_2)^2\right)\right]+2 [4 s_2+(m_1+m_2+s_2) (m_2+2 s_2)] s_3+s_2 s_3^2}{2 (8 s_2+s_3 (8+s_3))}\epsilon_2^2\nn\\
&&+\tfrac{2 s_2^2 s_3-2 m_1^2 (4+s_3)-4 m_1 [s_3+m_2 (2+s_3)-4]+s_2[8-8 m_2+3 s_3 (4+s_3)]+s_3 [-2 m_2 (6+m_2)+(3+s_3) (8+s_3)]}{8 s_2+s_3 (8+s_3)}\epsilon_1\epsilon_2\nn\\
&&+\tfrac{-32 \left[2 m_1^2+3 m_1 (m_2-2)+(m_2-2)^2+(m_2-1) s_2\right]+8 [4+m_2 (m_1+m_2-s_2-6)+3 s_2] s_3+4 (11-3 m_1-5 m_2+s_2) s_3^2+7 s_3^3}{8 s_2+s_3 (8+s_3)}\epsilon_1\nn\\
&& +\Big[\tfrac{-8 m_1 (m_2-2) (2+s_2)-8 s_2 [-2-s_2+m_2 (2+m_2+s_2)]-4 m_1 (2+4 m_2+s_2) s_3}{8 s_2+s_3 (8+s_3)}\nn\\
&&\qquad\qquad\qquad\qquad+\tfrac{
4 \left[-3 m_2^2-3 m_2 (2+s_2)+(2+s_2) (6+s_2)\right] s_3+(22+7 s_2) s_3^2+2 s_3^3-8 m_1^2 (4+s_3)}{8 s_2+s_3 (8+s_3)}\Big]\epsilon_2\nn\\
&&+ \tfrac{4 \left(-32 m_1^2+4 m_1 (2 m_2-s_3-4) (s_3-4)+(4-2 m_2+s_3)^2 (-2+s_2+2 s_3)\right)}{8 s_2+s_3 (8+s_3)}\Big\}~.
\eea
Extremizing this with respect to $\epsilon_1,\epsilon_2$ then gives the on-shell central charge. 
For example, setting the flavour twist parameters $\pt_2=0=\pt_3$, we obtain (again for $X^{2,1}$)
\bea
\csugra  \ = \  \frac{48 (g-1)N^2 m_1 \left[(2 m_1^2 +7m_1m_2)(m_2-1)+2 m_2^2 (3 m_2-4)+2m_1 \right] }{m_1 [m_1 (2+m_1)-12]+m_1 (4+3 m_1) m_2+(3 m_1-2) m_2^2+m_2^3}~.
\eea
It is straightforward to also compute the off-shell (and hence on-shell) $R$-charges $R_a$, $a=1,\ldots,5$, although we do not record the formulae here.

\subsubsection*{Dual field theory and $c$-extremization}

We next move on to the dual field theory computation, with the 
four-dimensional $X^{p,q}$ quiver gauge theories wrapped on $\Sigma_g$. 
The theory has $SU(N)^{2p+1}$ gauge group and the field content is summarized in
Table \ref{table2}.
\begin{table}[h!]
\small{
\begin{center}
\begin{tabular}{|c|c|c|c|c|c|c|}
\hline
Field&Multiplicity&$R_0$-charge&$U(1)_{B_{1}}$&$U(1)_{B_{2}}$&$U(1)_{F_{1}}$&$U(1)_{F_{2}}$\\
\hline\hline
$X_{12}$&$p N^2$             &$0$   &$0$      &$-1$         &$1$&0\\
$X_{23}$&$(p+q-1)N^2$   &$2$   &$1$      & $2$          &$0$&0\\
$X_{34}$&$N^2$                &$0$   &$-p$     &$-p-q$     &$0$&$1$\\
$X_{45}$&$N^2$                &$0$   &$p$      &$p+q-1$  &$0$&$-1$\\
$X_{51}$&$(p-q)N^2$        &$0$   &$-1$     &$0$          &$-1$&0\\
$X_{24}$&$N^2$                &$2$   &$1-p$   &$2-p-q$   &$0$&$1$\\
$X_{31}$&$(q-1)N^2$        &$0$   &$-1$     &$-1$          &$-1$&$0$\\
$X_{35}$&$(p-1)N^2$        &$0$   &$0$      &$-1$          &0&0\\
$X_{41}$&$N^2$                &$0$   &$p-1$   &$p+q-1$  &$-1$&$-1$\\
$X_{52}$&$q N^2$             &$0$   &$-1$     &$-1$         &0&0\\
$\lambda$&$(2p+1)(N^2-1)$&1&0&0&0&0\\
\hline
\end{tabular}
\caption{The field content of the $X^{p,q}$ quiver theories.}
\label{table2}
\end{center}
}
\end{table}
There are now two baryonic symmetries, $U(1)_{B_I}$, $I=1,2$,
associated to the two non-trivial three-cycles, and two (non $R$-symmetry) flavour symmetries
$U(1)_{F_i}$, $i=1,2$, corresponding to $U(1)$ isometries  under which the holomorphic volume 
form $\Omega_{(3,0)}$ is uncharged. Once again $R_0$ is a simple fiducial charge 
to be used in the $c$-extremization procedure.
The topological twist is along the generator
\bea\label{dp2top}
T_{\mathrm{top}} & =&f_1T_{F_1}+  f_2T_{F_2}+\betabeta_1 T_{B_1}+\betabeta_2 T_{B_2}+\frac{\kappa}{2}R_{R_0}~,
\eea
with the trial $R$-charge being
\bea\label{dP2trialR}
T_{\text{trial}} &=&  T_{R_0}+\epsilon_{B_1}T_{B_1}+\epsilon_{B_2}T_{B_2}+\epsilon_{1} T_{F_{1}}+\epsilon_{2} T_{F_{2}}~.
\eea

As in (\ref{cdP2grav}) we present the trial $c$-function only for the $X^{2,1}$ case:
\bea
&& c(\epsilon_{B_1},\epsilon_{B_2},\epsilon_1,\epsilon_2) \ =\  
3 N^2 \bar\eta  \Big\{2 f_2 \epsilon_1+f_2 \epsilon_1^2-8 B_1 \epsilon_2+4 f_2 \epsilon_2-2 B_1 \epsilon_1 \epsilon_2+2 f_2 \epsilon_1 \epsilon_2\nn\\
&& +16 B_1 \epsilon_{B_1} -8 f_2 \epsilon_{B_1}+4 B_1 \epsilon_1 \epsilon_{B_1}-2 f_2 \epsilon_1 \epsilon_{B_1}-2 B_2 [2 \epsilon_1 (2+\epsilon_2-\epsilon_{B_1})-8 \epsilon_{B_1}+\nn\\
&& + \epsilon_2 (6+\epsilon_{B_1})]+16 B_1 \epsilon_{B_2}-12 f_2 \epsilon_{B_2}+4 B_1 \epsilon_1 \epsilon_{B_2}-4 f_2 \epsilon_1 \epsilon_{B_2}+4 B_2 (6+\epsilon_1-\epsilon_2) \epsilon_{B_2}\nn\\
&& -2 B_1 \epsilon_2 \epsilon_{B_2}-2 f_2 \epsilon_{B_1} \epsilon_{B_2}-2 f_2 \epsilon_{B_2}^2+f_1 [\epsilon_2^2+2 \epsilon_1 (4+\epsilon_2)-8 \epsilon_{B_2}+2 (\epsilon_{B_1}+\epsilon_{B_2})^2\nn\\
&& -2 \epsilon_2 (-1+\epsilon_{B_1}+2 \epsilon_{B_2})]+[2 \epsilon_1^2+(\epsilon_2-2 \epsilon_{B_1})^2+\epsilon_1 (\epsilon_2-4 \epsilon_{B_2})-6 \epsilon_2 \epsilon_{B_2}\nn\\
&& +8 \epsilon_{B_1} \epsilon_{B_2}+6 \epsilon_{B_2}^2]) \kappa \Big\}~.
\eea
We next 
extremize over both of the baryon mixing parameters $\epsilon_{B_1}$, $\epsilon_{B_2}$, by solving
\bea\label{mixdP2}
\frac{\partial c}{\partial \epsilon_{B_1}} &=& 0 \ = \ \frac{\partial c}{\partial\epsilon_{B_2}}~,
\eea
for $\epsilon_{B_1},\epsilon_{B_2}$,
and then substitute back into the trial $c$-function to obtain $c=c(\epsilon_1,\epsilon_2)$.
For $|\kappa|=1$ we then find, remarkably,
this function exactly matches the off-shell trial central charge computed in gravity (\ref{cdP2grav}), after 
making the appropriate simple change of basis
\bea\label{cobdp2}
B_1&=&  \frac{1}{2} (2 - m_1 + \pt_2) \kappa~, \qquad B_2 = -\frac{1}{2} (2 + m_2 + \pt_2) \kappa~, \nn\\
\quad f_1 &=&  -\frac{1}{2} (2 + \pt_2) \kappa~, \qquad \qquad  f_2 = -\frac{1}{2}(2p + \pt_3) \kappa~.
\eea
Moreover, the trial $R$-charges $R_a$, $a=1,\ldots,5$, in  field theory and gravity also match off-shell, 
for general $m_1,m_2,\pt_2,\pt_3$, as a function of $\epsilon_1,\epsilon_2$, after resolving the baryon mixing 
(\ref{mixdP2}) in field theory.  Specifically
\bea\label{compareRsdp2}
R_a \ = \ R[X_a]N~, \qquad a=1,\ldots,5~,
\eea
where $(X_1,X_2,X_3,X_4,X_5)\equiv (X_{51},X_{12},X_{23},X_{34},X_{45})$, and the $X_{ij}$ are the 
bifundamental fields in the $X^{p,q}$ quiver, as listed in
table \ref{table2} above. Furthermore, using the formulae for these field
theory $R$-charges in \eqref{togetbeprel}, we obtain the change of variables given in \eqref{beetoepdp2}.

As we saw in the $Y^{p,q}$ examples, the simple change of basis given in \eqref{cobdp2} can again be obtained by solving
\begin{align}
M_a=-\bar\eta\, t[X_a] N\,,
\end{align}
where the $M_a$ are the fluxes on the gravity side, given in \eqref{mandnxpq}, \eqref{mandnxpq2}, 
and $t[X_a]$ are the charges of the bosonic fields $(X_1,X_2,X_3,X_4,X_5)\equiv (X_{51},X_{12},X_{23},X_{34},X_{45})$
in the quiver gauge theory
with respect to the background gauge field given in \eqref{dp2top}. 

\section{Explicit supergravity solutions for $Y^{p,q}$ case}
\label{secexplexamples}

In the previous section we matched the central charge and $R$-charges on the gravity side,
obtained using our new toric geometry formalism, with those from field theory using $c$-extremization. 
As anticipated in section \ref{sec:exist}, this is \emph{a priori} a formal matching. 

On the gravity side, we have shown how to calculate the
central charge and $R$-charges for a class of AdS$_3\times Y_7$ solutions
of type IIB supergravity, for $Y_7$ of the fibred form 
$Y_5\hookrightarrow Y_7\rightarrow  \Sigma_g$, using the toric data of $Y_5$,
provided that the solution actually exists. In particular, having computed the central charge 
and $R$-charges for particular toric data and twisting over $\Sigma_g$, one should then 
check that these quantities are positive. 
From the field theory side, the $c$-extremization
procedure will give the correct central charge and $R$-charges, provided that the putative SCFT field
theory, obtained from compactifying the quiver gauge theory with particular topological twists, actually exists. It is certainly possible that in some cases the compactified field theory flows to
some other behaviour in the IR. However, demanding positivity of the central charge and $R$-charges places strong necessary conditions for the existence of the SCFT fixed point, and it is natural
to anticipate that, generically, for the class of field theories we are studying, these also
provide sufficient conditions. 

An additional point to highlight on the gravity side is that our geometric results 
used the toric data for K\"ahler cones over $Y_5$, ${C}(Y_5)$,
specified by the inward pointing normal vectors $\v_a$ associated with the convex polyhedral cone. However, as discussed in \cite{Couzens:2018wnk}, we can also use our geometric results, at least formally, for complex cones ${C}(Y_5)$ with a holomorphic $(3,0)$-form and $U(1)^3$ action
which do not admit any compatible K\"ahler cone metric. In this case we can still define vectors
$\v_a\in \mathbb{Z}^3$, which define the $d$ complex codimension one submanifolds where
the action of one of the $U(1)\subset U(1)^3$ degenerates. These examples were called ``non-convex" toric cones in \cite{Couzens:2018wnk}, since the $\{\v_a\}$ do not define a convex polyhedral cone.
It would also be interesting to put our calculations on a firmer geometric footing for this class.

To illustrate several of these issues, we recall the example of the quiver gauge theory for the $Y^{p,q}$ 3-fold singularities compactified on $T^2$ with baryon flux only. From the gravity side, we discussed this example just below equation \eqref{Rn2} (it is the special case with $g=1$ and $n_2=0$). 
For toric $Y^{p,q}$, which
necessarily have $p>q>0$, this case is in fact obstructed, as proved in \cite{Couzens:2018wnk}. 
Specifically, the critical $R$-symmetry vector $\vec{b}$ lies outside the Reeb cone, {\it c.f.} the discussion in 
section \ref{sec:exist}.
However, if we consider non-convex toric cones with $q>p>0$ then, as we noted around equation
\eqref{emmsspc}, we obtain results using our toric formula which agree with the known explicit
supergravity solutions already constructed in \cite{Donos:2008ug}. 
On the field theory side, as already discussed in 
\cite{Couzens:2018wnk}, we see from \eqref{Rspecial} that the quiver gauge theories for 
the $Y^{p,q}$ 3-fold singularities, with $p>q>0$ cannot flow to a SCFT in the 
IR since the $R$-charges would be negative. Furthermore, this also shows
that there is no obvious candidate field theory that is dual to the explicit supergravity solutions
associated with the non-convex toric geometries with $q>p>0$.

The above discussion emphasizes that with our current understanding, it is illuminating to compare the formal matching we have demonstrated in this paper with explicit supergravity solutions. In section \ref{unitwist} we discussed the compactification of toric quiver gauge theories on a Riemann surface with genus $g>1$ and a universal twist. In this case we know that the explicit 
supergravity solution exists ({\it i.e.} they are not obstructed), and the agreement with $c$-extremization
in the field theory provides strong evidence that the field theory does indeed flow to a SCFT.
In the remainder of this section we will consider two further examples, for which an explicit supergravity
solution exists. The first is the $Y^{p,0}$ theories compactified on a Riemann surface with $g>1$, where one adds baryonic flux to the 
universal twist, for which an explicit supergravity solution was found in 
\cite{Gauntlett:2006qw, Donos:2008ug}, and we find a consistent picture analogous to the example of the universal twist. The second example we consider is associated with $Y^{p,q}$ fibred over $S^2$, and this example is analogous to the $T^2$ example discussed in the previous paragraph.
Applying $c$-extremization to the quiver gauge theory associated with $Y^{p,q}$ with $p>q>0$ does
not lead to physical results for the central charge and $R$-charges. On the other hand, there is a supergravity solution associated with a non-convex toric cone with $q>p>0$, whose central charge and $R$-charges agree with the calculations we obtained in \eqref{cchgegenypq} and
\eqref{rchgegenypq}.

In the two explicit AdS$_3\times Y_7$ solutions of the form \eqref{ansatz}
that we discuss, the central charge was 
calculated using the formula (equivalent to \eqref{cS2}, \eqref{cS23})
\bea
 \csugra & = &
 \frac{3L^8}{16\pi^6\ell_s^8 g_s^2}\int_{Y_7} \me^{-2B} \mathrm{vol}_7~,
 \label{explicitc}
\eea
where the volume form is with respect to $\dd s^2_{7}$ .
The $R$-charges were not calculated, but we do so here. When
$Y_7$ is a fibration of a $Y_5$ over a Riemann surface $\Sigma_g$, the 
holographic $R$-charges can be computed via the formula (equivalent to \eqref{rchgegenexpgeneral}) \cite{Couzens:2017nnr}
\bea
R_a \ = \ R[S_a] & = & \frac{L^4}{8 \pi^{3} \ell_s^4 g_s} \int_{S_a} \me^{-B}\mathrm{vol}(S_a)~,
\label{explicitrs}
\eea
where $S_a$ are supersymmetric three-manifolds in $Y_7$, and the volume form $\mathrm{vol}(S_a)$ is computed with respect to the pull-back of $\dd s^2_{7}$.

\subsection{$Y^{p,0}$ fibred over $\Sigma_{g>1}$}

In the AdS$_3\times Y_7$ solutions discussed\footnote{They were also discussed earlier in the context of AdS$_3$ solutions of 
$D=11$ supergravity: see section 6 of \cite{Gauntlett:2006qw} with $B_6=T^2\times KE_2^+\times KE_2^+$ and $c_1=0$.}
in section 3.1 of \cite{Donos:2008ug}, $Y_7$ has a six-dimensional transverse K\"ahler metric which consists of a 
product of three K\"ahler-Einstein metrics, $\Sigma_g\times S^2_1\times S^2_2$,  with the genus $g>1$.
At a fixed point on $\Sigma_g$ one finds a copy of $T^{1,1}=Y^{1,0}$, so that the total space is a fibration of $Y^{1,0}$ over $\Sigma_g$. A simple supersymmetry-preserving quotient by $\Z_p$ yields  the total space  \cite{Benini:2015bwz}
\bea
Y^{p,0}\ \hookrightarrow \ Y_7 \ \rightarrow \ \Sigma_g~.
\eea
In \cite{Benini:2015bwz} the solution was identified as a holographic dual to the $Y^{p,0}$ quiver theory compactified on  $\Sigma_g$ with a particular baryonic twist, and the gravitational central charge computed with (\ref{explicitc}) was shown to agree exactly with the extremized $c$-function in the field theory. 
Below, our aim will be to illustrate the agreement of the explicit solution
with results that we obtained in section \ref{secypqexamples}, valid for $Y^{p,q}\hookrightarrow Y_7 \rightarrow \Sigma_g$, with arbitrary   baryon and ($SU(2)$-preserving) flavour fluxes. We will additionally check that the
 $R$-charges, obtained in section \ref{secypqexamples}, match with the $R$-charges computed using (\ref{explicitrs}), as expected. We note that the solution is specified\footnote{One can check that the solution presented in section 6 of \cite{Donos:2008ug} and that presented in section 3.3 of \cite{Benini:2015bwz} coincide, upon identifying the parameters as $l_1=\frac{s}{t}=-\frac{1}{v+1}$, and $-\left(\frac{t}{h}N\right)_{\mathrm{DGK}}=N_{\mathrm{BBC}}\equiv N$.} 
 by a rational number $v\ge 1$ in addition to $p$.

To make the comparison, we have to give relations between the parameters $N,M_a,n_i$ used in section \ref{secypqdp2} to the integer fluxes $N_A$ in the explicit supergravity solution. This can be determined by scrutinizing the five-cycles used to perform flux quantization in \cite{Gauntlett:2006qw}. In particular, three natural five-cycles  $D_1,D_2,D_3$ were discussed, subject to the homology relation $[D_1]+[D_2]+(1-g)[D_3]=0$, corresponding to 
$U(1)$ fibrations over four-cycles $\Sigma_g\times S_1^2$, $\Sigma_g\times S_2^2$, and $S_1^2\times S_2^2$, respectively. 
Since the five-cycle obtained by fixing a point on $\Sigma_g$ is a copy of $Y^{p,0}$, the flux through this can be identified with the total number of D3-branes, that we denote by $N$, namely
\bea
N(D_3) & = & N~.
\eea
The two other fluxes are given by 
\bea
N(D_1) & = &  \frac{g-1}{v+1}N ~,\quad   N(D_2) ~ = ~ \frac{v(g-1)}{v+1}N~.
\eea
By examining the supergravity solution one can identify these fluxes with our $M_a$ 
via
\bea\label{relatefluxesBBC}
M_1 \ = \  M_3 \ = \ N(D_1)~,\qquad M_2 \ = \ M_4 \ = \ N (D_2)\,,
\eea
and inserting these into (\ref{Marel}) we further obtain 
\bea
n_1\  =\  2-2g  ~, \quad n_2~=~n_3~= \ p(1-g)~,
\eea
where we used the $Y^{p,q}$ toric data (\ref{Ypqv}), with $q=0$. 
Furthermore, from \eqref{Mandn},\eqref{Mandn2} we can make the identification
\bea\label{emexypz}
m &=& \frac{1}{1+v}~, \qquad s\ = \ -p~.
\eea 

Having made these identification we can now compare the supergravity central charge  and $R$-charges, computed using the explicit metric,
with the  results of the analysis in section \ref{secypqexamples}. 
After substituting \eqref{emexypz} into the general expression for the central charge (\ref{cchgegenypq}), we obtain
\bea
\csugra &=& 6p(g-1) \frac{v^2+v+1}{(v+1)^2}N^2~,
\label{fworksc}
\eea
in agreement with the result computed in \cite{Gauntlett:2006qw,Benini:2015bwz}. 
It is straightforward to compute the $R$-charges  
from the explicit solution using (\ref{explicitrs}), 
finding that they agree with those presented in  (\ref{rchgegenypq}) after substituting
\eqref{emexypz}, namely
\bea
R_1 &=& R_3 \ = \ \frac{v}{v+1}N~, \qquad R_2\ = \ R_4 \ = \ \frac{1}{v+1}N~.
\label{fworksrs}
\eea
Notice that for $v=1$ the solution coincides with the universal twist of  $Y^{p,0}$, and both  (\ref{fworksc}), (\ref{fworksrs}) reduce to their correct values. 

In section  \ref{secypqexamples} we also showed that the trial central charge and $R$-charges agree (even off-shell) with the corresponding quantities computed employing $c$-extremization.  In particular, with \eqref{emexypz}
the twisting parameters $B$ and $f_2$ given in (\ref{Bf2}) take the values 
$B=-\frac{1}{2p(v+1)}$ and $f_2=0$. This indicates that there are additional supergravity solutions
associated with the more general values of $B$ and $f_2$. Also,
recall that the baryon twisting parameter $B$ that we defined in section \ref{secypqexamples} does not coincide with the parameter ``$B$'' defined in \cite{Benini:2015bwz}. In particular, in the basis we have chosen the baryon twist parameter 
does not vanish for the universal twist $v=1$, while that in \cite{Benini:2015bwz} does vanish.

\subsection{$Y^{p,q}$ fibred over $\Sigma_{g=0}$}

\newcommand{\pJ}{\mathtt{p}}
\newcommand{\qJ}{\mathtt{q}}

The explicit AdS$_3\times Y_7$ solutions presented  in \cite{Gauntlett:2006af} have a $Y_7$ that is constructed\footnote{This construction is analogous to the one for Sasaki-Einstein seven-manifolds described in 
\cite{Gauntlett:2004hh,Martelli:2008rt}.} as a fibration over a positively curved K\"ahler-Einstein four-manifold.  In particular, taking the $KE_4$
to be $S^2\times S^2$, the total space $Y_7$ is toric, and it  may also be viewed as the fibration of a  toric $Y_5$ over either of the two Riemann spheres, $\Sigma_0=S^2$. The fibre manifold, $Y_5$, at a fixed point on the two-sphere base, $\Sigma_0$, has topology $S^2\times S^3$, isometry group $SU(2)\times U(1)^2$ and is labelled by two coprime integers $\pJ,\qJ>0$ (which were labelled $p,q$ in \cite{Gauntlett:2006af}). 

For $KE_4 = S^2\times S^2$ the integers  $(M,m)$ in 
\cite{Gauntlett:2006af} take the values $M=8$, $m=2$. 
Now, if we fix a point on $\Sigma_0=S^2$ in   \cite{Gauntlett:2006af}, the topological construction of the metrics in \cite{Gauntlett:2006af} is very similar to that in \cite{Gauntlett:2004yd}, 
with a simple re-labelling of the parameters. 
 In particular the  $(2\pi)$-periodic  coordinate $\psi$ in the two constructions is the same and together with the coordinate $y$ these form a (topologically trivial) $S^2$ bundle over $S^2$, that was denoted $B_4$ in \cite{Gauntlett:2004yd}. 
 One then constructs a circle bundle over this four-dimensional base,
with Chern numbers  (in the notation of \cite{Gauntlett:2004yd}) $p$ and $q$. In particular, $p$ is the Chern number over the fibre $S^2$. 
 From equation (12) of  \cite{Gauntlett:2006af} we thus immediately read off
\bea\label{qredef}
\qJ &=& p~.
\eea
A careful comparison\footnote{Recall that in the $Y^{p,q}$ construction of \cite{Gauntlett:2004yd}, the four-dimensional base $B_4$ is topologically 
$S^2\times S^2$, with generating two-cycles $C_1$ and $C_2$. These are related to the north and south pole sections $S_1$ and $S_2$ 
of the fibre $S^2$ via $2C_1=S_1-S_2$, $2C_2=S_1+S_2$. 
One easily checks that the Chern numbers over $S_1$ and $S_2$ are $p+q>0$ and $-p+q<0$. This then matches 
with the Chern numbers $p$ and $q$ over $C_1$ and $C_2$, using the above relation between cycles. See also \cite{Couzens:2017nnr} for a similar comparison of parameters between different versions of $Y^{p,q}$ manifolds.} 
 of the basis of two-cycles $C_1$ and $C_2$ in $B_4\simeq S^2 \times S^2$ (which should not be confused with the $KE_4 = S^2\times S^2$) 
leads to identify the parameter $\pJ$ in \cite{Gauntlett:2006af} with 
\bea\label{predef}
\pJ &=& -p+q~.
\eea
 Note that since $\pJ>0$, $\qJ>0$ in the supergravity solutions of 
 \cite{Gauntlett:2004yd}, we now crucially have $q>p$, which is the opposite inequality to the 
Sasaki-Einstein $Y^{p,q}$ metrics. Thus, similar to the case discussed in \cite{Couzens:2018wnk}, 
we cannot strictly compare with our general toric formalism, since the 
toric data (\ref{Ypqv}) when $q>p$, that are relevant for the supergravity solutions of
 \cite{Gauntlett:2006af}, do not form a convex set. However, we can formally apply 
our toric formulas and, remarkably, we find precise agreement with the supergravity results. 

To make the comparison we need to relate the parameters $N$, $M_a$, $n_i$ used in section~\ref{secypqdp2} to the integer fluxes $N_A$ in the supergravity solution of \cite{Gauntlett:2006af}. Now the supergravity solutions have $H_5(Y_7;\Z)=2$, 
but the quantization conditions in \cite{Gauntlett:2006af} were described in terms of four five-cycles $D_0,\tilde D_0, D_1,D_2$ subject to linear relations. Here $D_0$  and $\tilde D_0$ denote the five-cycles arising from two sections of the fibration over $KE_4 = S^2\times S^2$, while $D_1, D_2$ are the five-cycles corresponding to the fibration over the generating two-cycles in $KE_4 = S^2\times S^2$. 

Since the five-cycle obtained by fixing a point on $\Sigma_0=S^2$ is a copy of $Y^{p,q}$ (with $q>p>0$) 
the associated flux through this cycle can be identified with the total number of D3-branes, $N$. 
Therefore, from (18) and (19) of \cite{Gauntlett:2006af} we have\footnote{Here we have also set the two integers $n_a$ of \cite{Gauntlett:2006af} to unity. This follows by noting that since
the canonical bundle of $S^2\times S^2$ is $O(-2,-2)$ and $m=2$, the line bundle $\mathcal{N}$
of \cite{Gauntlett:2006af} is $O(-1,-1)$ and hence $n_a=1$.}, in the notation here ({\it i.e.} after using
\eqref{qredef}, \eqref{predef}),
\bea
N(D_1)\  =\  N (D_2) \ = \  \frac{n}{h} p  \  =  \ N~,
\eea
while the two remaining fluxes given in (18) of \cite{Gauntlett:2006af}, in our notation, become
\bea
N(D_0) \ = \  -\frac{2N}{p}(p+q)~,\quad 
N(\tilde{D}_0) \  = \     -\frac{2N}{p}(-p+q)~.
\eea
Further examination of the supergravity solution allows us to
identify these fluxes with our $M_a$ via
\bea\label{emmssec7}
M_1 ~= ~ N(D_0)~,\quad  M_2 ~=~M_4~= ~N ~,\quad    M_3 ~= ~ - N( \tilde D_0)  ~.
\eea
Inserting these into (\ref{Marel}) we also obtain 
\bea
n_1 \ = \ 2-2g \ = \ 2 ~, \quad n_2~=~n_3~\ =\ p-q~,
\eea
and comparing 
\eqref{emmssec7} with \eqref{Mandn}, \eqref{Mandn2} implies that
\bea\label{emess}
m ~=~ \frac{2}{p}(p+q)~,\qquad s~=~q-p~.
\eea

We can now compare the supergravity central charge  
and $R$-charges computed using the explicit metric with the
results of the analysis in section \ref{secypqexamples}.
The central charge for the explicit supergravity solutions given in eq. (1) of \cite{Gauntlett:2006af}
reads, in the notation here, 
\begin{align}
\csugra\ =\ \frac{18 p (q-p) (p+q)N^2}{p^2+3 q^2}~.
\end{align}
One can check that this precisely agrees with the toric calculation (\ref{cchgegenypq}) after using
\eqref{emess}.
It is also straightforward to compute the $R$-charges (\ref{explicitrs}) from the explicit supergravity solution (which was not done in \cite{Gauntlett:2006af}),
finding that they also agree with those presented in  (\ref{rchgegenypq}), namely
\begin{align}
R_1 &\ =\ \frac{ (3q-p) (p+q)N}{p^2+3 q^2}\,,\nonumber\\[2mm]
R_2 ~= ~R_4 &\ =\ \frac{2  p^2N }{p^2+3 q^2}\,,\nonumber\\[2mm]
R_3 &\ =\ \frac{ (q-p) (p+3 q)N}{p^2+3 q^2}~.
\end{align}
Notice that $\csugra$ and $R_3$ are positive if and only if $q>p$, associated with the explicit
supergravity solutions. As we remarked earlier, the derivation of our formulas assumed that the toric data of the $Y_5$ fibred over $S^2$ forms a convex set, which does not hold when $q>p$. Nevertheless, we find perfect agreement with the explicit metric in this case. Something similar was also seen for different examples in
\cite{Couzens:2018wnk}, which strongly suggests that our general toric formulas continue to
hold outside the regime of validity of their derivation. It would be interesting to explore this further. 

In section  \ref{secypqexamples} we also showed that the trial central charge and $R$-charges agree (even off-shell) with the corresponding quantities computed employing $c$-extremization in the 
dual four-dimensional quiver gauge theories reduced on $\Sigma_0$. In fact, this matching holds also by formally taking $q>p$ in the trial $c$-function of the field theory. In this case, with
\eqref{emess},
the twisting parameters $B$ and $f_2$ given in (\ref{Bf2}) take the values 
$B=\frac{1}{p}$ and $f_2=0$. Since in the field theory there is no reason to assume the baryon and flavour fluxes to be fixed to any particular values, we are led 
to conjecture that there exist supergravity solutions generalizing the one we discussed here,  corresponding to $Y^{p,q}$ fibred over $\Sigma_0=S^2$ with arbitrary twisting parameters $m$ and $s$ (equivalently arbitrary values of 
$B$ and $f_2$, consistent with  (\ref{Bf2})).

\section{Discussion}\label{sec:discussion}

In this paper we have elaborated on the  extremal problem  recently formulated in \cite{Couzens:2018wnk}, which was proposed as a geometric 
dual to the  procedure of $c$-extremization \cite{Benini:2012cz,Benini:2013cda} for two-dimensional $(0,2)$ SCFTs.  
By analogy with the geometric dual to $a$-maximization \cite{Intriligator:2003jj} put forward in
\cite{Martelli:2005tp}, it was shown  in \cite{Couzens:2018wnk}
that the $R$-symmetry Killing vector field characterizing the class of odd-dimensional 
``GK geometries" $Y_{2n+1}$ considered in \cite{Gauntlett:2007ts} may be determined by extremizing
a function that depends only on certain global, topological data, subject to some constraints. In dimensions $n=3$ and $n=4$, these geometries arise in the context of supersymmetric AdS$_3\times Y_7$ solutions of type IIB supergravity, and 
AdS$_2\times Y_9$ solutions of eleven-dimensional supergravity, respectively. In these cases, the extremal problem 
determines necessary conditions to solve  a curvature condition  arising from the supergravity equations of motion, and the constraints are determined 
by the Dirac quantization conditions  of the fluxes through  cycles in $Y_{2n+1}$. 
 
Focusing on the case $n=3$, in this paper we developed a formalism that allows one to efficiently compute all the quantities necessary for implementing the extremization problem, when the seven-dimensional manifold $Y_7$ is a fibration of a \emph{toric} $Y_5$ over a Riemann surface $\Sigma_g$ of genus $g$. As we have explained, the formulas that we derived are direct extensions of the corresponding expressions for toric Sasakian manifolds, presented in \cite{Martelli:2005tp}. Similarly to \cite{Martelli:2005tp}, these  formulas allow one to extract considerable information about the solutions, without their explicit knowledge. 
We have illustrated in a number of explicit examples that the geometric quantities calculated using our formalism agree spectacularly with the corresponding quantities computed implementing $c$-extremization in the dual field theories, or extracted 
from the explicit supergravity solutions, when these are available. 

The results presented here open the way to a number of interesting research directions. An obvious extension is to generalize our formalism to toric geometries associated with dimension $n>3$. We expect that all the information necessary to implement the relevant extremal problem can still be encoded in a master volume $\mathcal{V}$, depending on the toric data 
$\v_a\in \Z^n$, the $R$-symmetry vector $\vec{b}\in \R^n$, and the K\"ahler parameters $\lambda_a$,  with the key formulas  (\ref{susyactform})--(\ref{Maformulaintro}) generalizing straightforwardly. 
Such an extension will allow one to study generic AdS$_3\times Y_7$ solutions of type IIB supergravity with $Y_7$ toric,
which may provide important clues to identifying the dual SCFTs, including for the known supergravity solutions \cite{Gauntlett:2006af,Gauntlett:2006qw,Gauntlett:2006ns}. In addition, such an extension can be used to study
AdS$_2\times Y_9$ solutions of eleven-dimensional supergravity, where $Y_9$ is toric or when
$Y_9$ is a fibration of toric $Y_7$ over a Riemann surface $\Sigma_g$. This latter case is
particularly interesting as it corresponds
to taking the $\mathcal{N}=2$, $d=3$ SCFTs dual to AdS$_4\times Y_7$ solutions,
arising from placing M2-branes at the singular apex of a toric Calabi-Yau 4-fold cone, and then
reducing them 
on $\Sigma_g$. Some explicit supergravity solutions of this type were studied in 
\cite{Azzurli:2017kxo}. 
From a technical point of view, a simpler and more immediate generalisation of the results of this paper is to consider $Y_{2n+1}$ GK geometries with $n>3$ which are fibrations of a toric $Y_5$
manifold over a K\"ahler-Einstein manifold of dimension $2n-4$. This would include 
AdS$_2\times Y_9$ solutions of $D=11$ supergravity with $Y_9$ obtained as a toric $Y_5$ fibred over a $KE_4$ manifold.

Here we have provided some necessary conditions for the existence of GK metrics, which have transverse K\"ahler geometries obeying the prescribed curvature condition (\ref{boxReqn}). 
While there are several infinite classes of explicit metrics that have been constructed 
\cite{Maldacena:2000mw,Gauntlett:2001qs,Gauntlett:2006af,Gauntlett:2006qw,Gauntlett:2006ns,Gauntlett:2007ts,Donos:2008ug,Donos:2011pn,Benini:2015bwz,Azzurli:2017kxo},
it is manifestly clear that this direct approach is limited. As a simple example, it is very unlikely that
explicit metrics could ever be constructed analytically for the examples discussed in section \ref{sec:dp2example}.
Thus, as discussed in section~\ref{sec:exist}, it is important to extend our work in the direction of establishing sufficient conditions for the existence of GK metrics, in the toric setting, analogous to the results for toric Sasaki-Einstein metrics proved in \cite{Futaki:2006cc}.
A remarkable fact that has emerged from our work is that 
while our master formula for the action is homogeneous of degree minus one in $\vec{b}$, after implementing the constraints  it somewhat miraculously becomes a quadratic function of the two remaining degrees of freedom. 
This implies that the critical $R$-symmetry vector is rational, namely $Y_7$ are quasi-regular, and  matches exactly the general expectation of the field theory analysis. \emph{We conjecture that $Y_7$ are always quasi-regular, including when they are not toric}.  We hope that some of these challenges will be taken up by the mathematics community. 

In all of the examples that we have analysed, the trial central charge function $\cZ$   associated to a (toric) geometry and the trial $c$-function of the dual two-dimensional $(0,2)$ SCFT have been shown to agree off-shell. More specifically, in all the examples that we have considered,
in which the field theories arise from compactifying four-dimensional SCFTs on a Riemann surface, the trial $c$-function depends on twisting parameters and trial mixing parameters for all the global (abelian) symmetries, both flavour and baryonic. On the other hand, from the geometric perspective, the function $\cZ$   depends on the same number of twisting parameters, related to quantized fluxes, but it has to be extremized only over the the flavour mixing parameters. \emph{We conjecture that after extremizing over the baryonic directions, the trial $c$-function in the field theory will match off-shell with the function $\cZ$.} 
We expect that this should not be difficult to prove, using ideas similar to  \cite{Butti:2005vn} (see also \cite{Lee:2006ru,Amariti:2017iuz}). 
In particular, we have seen in examples that the dictionary between the flavour trial parameters
 $\epsilon_1,\epsilon_2$ in the field theory and the free components of the $R$-symmetry vector $b_2, b_3$ on the gravity side
 can be implemented through the relation
\begin{align}\label{simprelMteedisc}
b_i&\ =\ \sum_{a=1}^d v_a^i R[X_a] ~,
\end{align}
where $R[X_a]$ are the $R$-charges in the field theory evaluated at the critical values of the field theory baryonic parameters.
Furthermore, the dictionary between the field theory baryonic and flavour twists $B_I$ and $f_i$, respectively, can be related to the independent subset of the fluxes $M_a$ and the free geometric twists $n_i$ on the gravity side via  
\begin{align}
M_a&\ =\ -\bar\eta\, t[X_a] N\,,
\end{align}
where $\bar\eta$ is given in \eqref{etadef2} and $t[X_a]$ are the charges of the basic fields in the quiver 
with respect to the topological twist in the field theory. Using \eqref{Marel} an immediate corollary is that 
\begin{align}
\bar\eta\sum_av^i_a t[X_a]=n_i\,.
\end{align}
 
One of the interesting novelties in our $\cZ$-extremization, compared to the results for Sasakian geometry \cite{Martelli:2005tp}, is the appearance of the 
K\"ahler parameters $\lambda_a$ in the problem. Generically these are eliminated in terms of the five-form flux quantum numbers $M_a$, which 
specify the dual $(0,2)$ theory. Specifically, they are related to the baryonic twist parameters $B_I$. 
On the other hand, in the field theory $c$-extremization the trial central charge is also a function of baryon mixing parameters $\epsilon_{B_I}$. In our geometric formulation there is no analogue of these variables $\epsilon_{B_I}$, and this is why we must extremize 
over these variables in field theory before matching to an off-shell $\cZ$ in gravity. However, this is perhaps suggestive  that the geometric extremal problem could be performed in an enlarged space. The natural setting in which this formulation could arise is that of the ``master space'' of  four-dimensional quiver gauge theories \cite{Forcella:2008bb}. If this is possible, one can also anticipate, for example, that it will be possible to derive the master volume, that we discussed in this paper, from a limit of an index-character defined on the master space, extending the known results in the Sasakian setting  \cite{Martelli:2006yb}.

\subsection*{Acknowledgments}
We would like to thank Chris Couzens for discussions.
JPG is supported by the European Research Council under the European Union's Seventh Framework Programme (FP7/2007-2013), ERC Grant agreement ADG 339140. JPG is also supported by STFC grant ST/P000762/1, EPSRC grant EP/K034456/1, as a KIAS Scholar and as a Visiting Fellow at the Perimeter Institute. JPG acknowledges hospitality from the KITP and
support from National Science Foundation under Grant No. NSF PHY-1748958.
DM is supported by the ERC Starting Grant 304806 ``The gauge/gravity duality and geometry in string theory''.




\providecommand{\href}[2]{#2}\begingroup\raggedright\endgroup

\end{document}